\documentclass[twocolumn]{aastex631}
\usepackage{amsmath}

\begin{document}

\title{Dynamical properties of magnetized low angular momentum accretion flow around a Kerr black hole}

\correspondingauthor{Indu K. Dihingia}
\email{ikd4638@sjtu.edu.cn, ikd4638@gmail.com}

\author[0000-0002-4064-0446]{Indu K. Dihingia}
\affiliation{Tsung-Dao Lee Institute, Shanghai Jiao-Tong University, Shanghai, 520 Shengrong Road, 201210, People's Republic of China}

\author[0000-0002-8131-6730]{Yosuke Mizuno}
\affiliation{Tsung-Dao Lee Institute, Shanghai Jiao-Tong University, Shanghai, 520 Shengrong Road, 201210, People's Republic of China}
\affiliation{School of Physics \& Astronomy, Shanghai Jiao-Tong University, Shanghai, 800 Dongchuan Road, 200240, People's Republic of China}
\affiliation{Institut f\"{u}r Theoretische Physik, Goethe Universit\"{a}t, Max-von-Laue-Str. 1, 60438 Frankfurt am Main, Germany}
 
\begin{abstract}
An essential factor in determining the flow characteristics of an accretion flow is its angular momentum. According to the angular momentum of the flow, semi-analytical analysis suggests various types of accretion solutions. It is critical to test it with numerical simulations using the most advanced framework available (general relativistic magnetohydrodynamics) to understand how flow changes with different angular momentum. By changing the initial condition of the accretion torus minimally, we can simulate steady, low angular momentum accretion flow around a Kerr black hole. We focus primarily on the lower limits of angular momentum and come upon that accretion flow with an intermediate range of angular momentum differs significantly from high or very low angular momentum flow. The intermediate angular momentum accretion flow has the highest density, pressure, and temperature near the black hole, making it easier to observe. We find that the density and pressure have power-law scalings $\rho\propto r^{n-3/2}$ and $p_g\propto r^{n-5/2}$ which only hold for very low angular momentum cases. With the increase in flow angular momentum, it develops a non-axisymmetric nature. In this case, simple self-similarity does not hold. We also find that the sonic surface moves away from the innermost stable circular orbit as its angular momentum decreases. Finally, we emphasize that intermediate angular momentum flow could provide a possible solution to explain the complex observation features of the supermassive black hole Sgr~A$^*$ at our galactic center.
\end{abstract}


\section{Introduction} \label{sec:intro}
Most astounding astrophysical events in the universe are driven by accretion onto compact objects like black holes and neutron stars \citep{Frank-etal2002}. Depending on the observational requirements, different kinds of astrophysical accretion flows have been proposed (e.g., spherically symmetric Bondi flow \citep{Bondi1952}, geometrically-thin accretion disk \citep{Shakura-Sunyaev1973, Novikov-Thorne1973}, advection dominated accretion flow (ADAF) \citep{Narayan-Yi1995}, multi-transonic accretion flows \citep{Fukue1987, Chakrabarti1989}, super-Eddington accretion flow \citep{Abramowicz-etal1988}, etc.) With the most advanced general-relativistic magneto-hydrodynamic (GRMHD) simulations, large numbers of studies have been performed to understand different kinds of flows \citep[See for review][]{Davis-Tchekhovskoy2020, Mizuno2022}. In the last few decades, some efforts have been put to understand low-angular accretion flow in pseudo-Newtonian, axisymmetric (2D), hydrodynamic approach targeting formation of standing as well as oscillating shocks \citep[e.g.,][]{Molteni-etal1994, Ryu-etal1995, Molteni-etal1996a, Molteni-etal1996b, Lanzafame-etal1998, Proga-Begelman2003, Chakrabarti-etal2004, Giri-etal2010, Okuda-Molteni2012, Okuda2014, Okuda-Das2015, Okuda-etal2019, Singh-etal2021, Okuda-etal2022}. However, so far, not much effort has been given to understanding low angular momentum flow in GRMHD framework which is related to multi-transonic accretion flows. 

Initially, \cite{Hawley-etal1984a, Hawley-etal1984b} developed a general relativistic hydro-dynamic (GRHD) framework to study relativistic 2D accretion flow around black holes. They have shown that the pressure-supported torus does not form for flow with angular momentum lower than that of marginally stable angular momentum. Recently, several authors have shown the possibility of multi-transonic flows and the presence of standing/oscillatory shocks in 2D GRHD accretion flow \citep{Sukova-etal2017, Kim-etal2017, Kim-etal2019, Palit-etal2019}. In three-dimensional (3D) GRHD simulations, \cite{Sukova-etal2017} show the formation of oscillatory and expanding shocks. However, recent 3D GRHD simulations do not show the signature of such standing shocks \cite{Olivares-etal2023}. Nonetheless, \cite{Olivares-etal2023} presented some local density jumps related to shocks in their simulations, which do not appear in global pictures. These results suggest that there are still some unexplored regions of the parameter space, and a systematic study of low angular momentum flow is needed. This study focuses on studying the accretion flow at different angular momentum limits to see how flow properties change with them.

All of the models in the simulation library for Sgr~A$^*$ and most of the models used for M~87$^*$ use variations of the same initial conditions of a rotation-supported torus seeded with a weak poloidal magnetic field \citep[][hearafter, FM torus]{Fishbone-Moncrief1976}. Undoubtedly, these models successfully constrain the parameters of these supermassive black holes, such as black hole mass, accretion rate, inclination, and black hole spin \citep{EHTI-2019, EHTVII-etal2021, EHT2022, EHT2024}. However, initializing simulations with a finite amount of matter in the torus leads to a drop in matter content and mass-accretion rate over time. Moreover, a stable FM torus solution does not exist for low angular momentum flow. In such flow (even in hydrodynamics), the gas pressure and lower centrifugal force are not enough to hold the torus in a stable structure against gravity. 
Accordingly, in this study, we employ the FM torus setup by changing the parameters from the ones used in earlier GRMHD simulations to achieve a steady state in low angular momentum accretion flows.

There is one more important piece of the puzzle: the GRMHD simulations show highly variable light curves than the observed light curve for Sgr~A$^*$ \citep{Murchikova-Witzel2021,EHT2022,Wielgus-etal2022,Murchikova-etal2022}. 
Although it is still not fully explained, so-called wind-fed accretion models are somehow preferred for accretion flow around Sgr~A$^*$ \citep[e.g.,][]{Murchikova-etal2022,Ressler-etal2023}. 
It is widely accepted that the winds of approximately thirty massive stars orbiting on the parsec scale may fuel in Sgr~A$^*$ \citep{Quataert2004,Cuadra-etal2008,Ressler-etal2018}.
Such models are not rotation-supported and have low angular momentum \citep{Ressler-etal2018}. Additionally, the hot-spots/flaring study of Sgr~A$^*$ suggests a magnetically arrested disk (MAD) around \citep[e.g.,] []{Dexter-etal2020,Porth-etal2021,Scepi-etal2022}. MAD around a rotating black hole is usually associated with a strong jet \citep[e.g.,][]{Tchekhovskoy-etal2011}.
On the contrary, there is no direct evidence for a jet in Sgr~A$^*$; morphological and kinematical studies suggest outflowing material or a weak jet from Sgr~A$^*$ \citep[see for discussion][]{Royster-etal2019,Yusef-Zadeh-etal2020}. With all these enigmas, it is worth exploring low angular momentum accretion flows and testing their relevance for Sgr~A$^*$.

In section \ref{sec:math}, we briefly discuss the numerical setup, and in the following sections (\ref{sec:result1},\ref{sec:result2},\ref{sec:result3}), we elaborately discuss the results obtained from our simulations models. Finally, in section \ref{sec:summary}, we summarise and discuss perspectives of our results. 

\section{Numerical setup}\label{sec:math}
This work investigates low angular momentum accretion flow with a set of three-dimensional (3D) ideal GRMHD simulations using the GRMHD code \texttt{BHAC} \citep{Porth-etal2017, Olivares-etal2019} in Modified Kerr-Schild (MKS) coordinates. The \texttt{code BHAC} assumes static spacetime, which means the mass of the central black hole is much larger than that of the surrounding matter. Thus, we neglect the self-gravity of the accretion flow in the simulations. We utilize a spherical polar grid denoted by $(r, \theta, \phi)$ with a logarithmic grid spacing in the radial direction ($r$, from $90\%$ of event horizon up to $r=2500\,r_g$) and a linear spacing in the polar ($\theta$) as azimuth ($\phi$) directions. The simulations are conducted in a generalized unit system where $G=M_{\rm BH}=c=1$. Here, $G$, $M_{\rm BH}$, and $c$ represent the universal gravitational constant, the mass of the black hole, and the speed of light, respectively. Distances and times are expressed in units of $r_g=GM_{\rm BH}/c^2$ and $t_g=GM_{\rm BH}/c^3$, respectively. To perform the simulations, we consider an effective resolution ($320\times128\times128$, with two static mesh refinement levels), where maximum resolutions are concentrated within $\pm45^\circ$ from the equatorial plane with $r\le100\,r_g$. With this, the minimum uniform grid size (in MKS, 2 levels) in the radial direction is $\Delta s_{\rm min} =0.02392$ and the maximum is $\Delta s_{\rm max}=0.04784$. In linear scale, the minimum radial grid size is $\Delta r_{\rm min}\sim 0.028692~r_g$ at the inner edge. The grid size increases with the increasing radius. For example at $r=100r_g$, the radial grid size is $\Delta r\sim 2.42072~r_g$.
For the sake of generalization, we consider a spinning black hole with Kerr parameter $a=0.9375$.
All simulations are performed up to $t=10,000\, t_g$.

\begin{figure}
    \centering
    \includegraphics[width=0.70\linewidth]{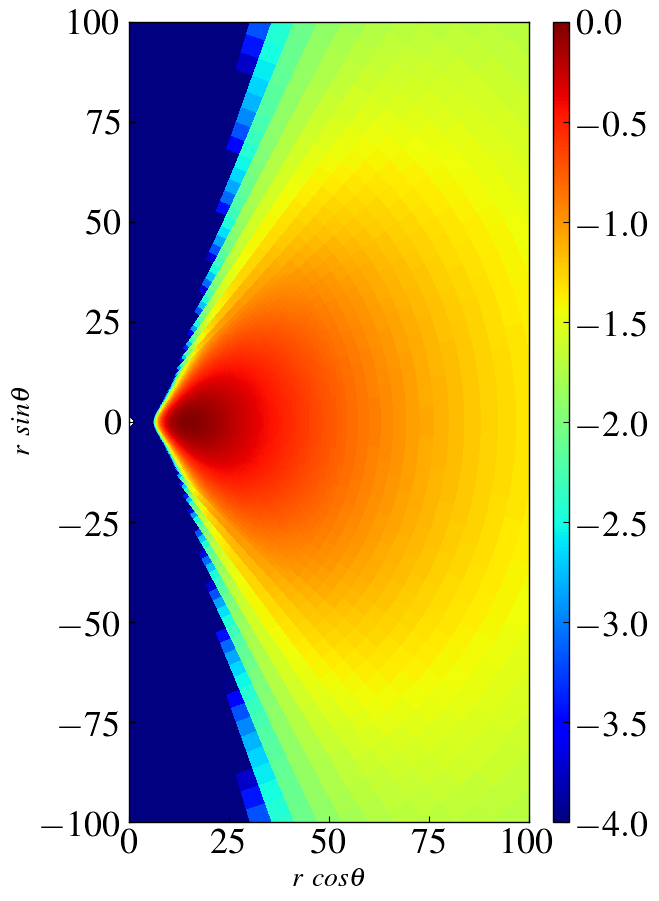}
    \caption{Initial logarithmic density distribution of the simulation models.}
    \label{fig:ini-den}
\end{figure}

This study does not intend to come up with a new, complicated way to simulate low angular momentum flows. For example, the wind-fed model \citep{Ressler-etal2018} or the perturbative simulation model \citep{Olivares-etal2023} are already being used for simulating low angular momentum flows around a black hole. We want to keep the spirit of the widely adopted, rotation-supported FM torus \citep{Fishbone-Moncrief1976} and modify the parameters minimally so that we can achieve steady-state solutions even for the lowest angular momentum. Accordingly, we set the initial density considering the FM torus distribution, with the inner edge at $r_{\rm min}=6\,r_g$ and the density maximum at $r_{\rm max}=15\,r_g$. These combinations give a density distribution with $\rho/\rho_{\rm max}\rightarrow6.626\times10^{-4}$ for $r\rightarrow\infty$ at the equatorial plane. Therefore, we have enough mass within the computational domain ($r<2500\,r_g$) to reach a quasi-steady state. Note that, for the usual torus setup (e.g., $r_{\rm min}=6\,r_g$, $r_{\rm max}=12\,r_g$, or $r_{\rm min}=20\,r_g$, $r_{\rm max}=40\,r_g$) with finite volume, the mass in the torus drained out before reaching a quasi-steady state for low angular momentum cases. The initial density distribution for the simulation models is shown in Fig.~\ref{fig:ini-den}. The figure shows that we allow accretion from a much wider direction as compared to the usual torus solutions.
Additionally, we supply angular momentum ($\lambda_0$) of the flow as a fraction of Keplerian angular momentum (${\cal F}$) at the $r_{\rm max}$ position, where $\lambda_0$ is the maximum angular momentum of the flow (see Appendix~A for detail profile).  In most cases, we choose the angular momentum to be less than the marginally stable angular momentum of the flow for the given Kerr parameter, i.e., $\lambda_{\rm ms}=2.3752$. Here we would like to mention that the marginally bound angular momentum for the given Kerr parameter is $\lambda_{\rm mb}=2.5$. We expect that, depending on the given angular momentum, the initial density at Fig.~\ref{fig:ini-den} will be redistributed and reach a quasi-steady accretion state after time evolution. Further, depending on the transport of angular momentum, we expect to have a distribution of angular momentum at the quasi-steady state for the given $\lambda_0$.
To study the impact of the angular momentum on the magnetic field configuration, we supply a poloidal single-loop magnetic field considering a non-zero azimuthal component of the vector potential, which is given by:
\begin{equation}
    A_\phi \propto \max(\rho/\rho_{\rm max}-10^{-3},0).
    \label{eq-aphi}
\end{equation}
To set the strength of the magnetic field, we choose the minimum ratio between gas pressure and magnetic pressure (plasma-$\beta$) to be $\beta_{\rm min}=100$. Following our motivations, we devised some simulation models, which we describe in Table~\ref{tab-01}. The radial profiles of other relevant quantities are shown in Appendix~A. Here, we choose one model without any magnetic field for comparison (\texttt{MOD5}). Due to the presence of a very high volume of matter within the simulation domain, we could reach a quasi-steady state even for the lowest angular momentum case (\texttt{MOD1}). However, if we keep on evolving them for a longer time, we expect to see a lowering and vanishing of the accretion rate at the event horizon. This is because we do not supply any matter from outside. Catch on to the fact that this property may look like runaway instability \citep{Abramowicz-etal1983}, but it is not the case as we neglect self-gravity in this work, which is crucial to excite runaway instability \citep[e.g.,][]{Masuda-Eriguchi1997,Korobkin-etal2013}. The goal of current work is not to explore these properties, and therefore, we do not evolve them further. Additionally, \texttt{BHAC} code is under development to consider self-gravity; we may be able to test such instability with the current framework in the future.
Note that model \texttt{MOD6} is a usual high-angular momentum torus with ${\cal F}=1$. 
\begin{table}
\centering
  \begin{tabular}{| l | c | c | c | c|}
    \hline
    Model & ${\cal F}$ &$\lambda_0$ & $\%$ of $\lambda_{\rm ms}$ & Mag. field \\ 
    \hline
    \texttt{MOD1} &  $0.2$ & $0.92$ & 38.7 & Yes \\
    \texttt{MOD2} &  $0.3$ & $1.38$ & 58.0 & Yes \\
    \texttt{MOD3} &  $0.4$ & $1.84$ & 77.4 &  Yes \\
    \texttt{MOD4} &  $0.5$ & $2.30$ & 96.7 &  Yes \\
    \texttt{MOD5} &  $0.5$ & $2.30$ & 96.7 &  No  \\
    \texttt{MOD6} &  $1.0$ & $4.60$ & 193.4 &  Yes \\
    \hline
  \end{tabular}
\caption{The explicit values of angular momentum fraction ${\cal F}$, specific angular momentum ($\lambda_0$), corresponding percentage of $\lambda_{\rm ms}$, and magnetic field status for different models are displayed.}
\label{tab-01}
\end{table}

In these simulations, we use a piecewise parabolic reconstruction scheme, the TVD Lax-Friedrichs approximate Riemann solver, an upwind constrained transport scheme to preserve the divergence-free constraint of the magnetic field, and 2nd-order Runge-Kutta time-stepping schemes \citep[][for detail]{Porth-etal2017,Olivares-etal2019}. Here, we reiterate the boundary conditions explicitly. We prohibit inflow at the inner radial boundary. Scalar and radial vector components are symmetric at the polar axis, while azimuthal and polar vector components are anti-symmetric. Additionally, imposed periodic boundary conditions along $\phi$ for all quantities. Our simulation setup does not allow matter inflow at the outer edge of the simulation domain. However, due larger simulation domain and modified initial density distribution, our simulations could reach a steady state to study accretion flow properties around black holes.

\section{Time series analysis}\label{sec:result1}
\begin{figure}
    \centering
    \includegraphics[width=1.0\linewidth]{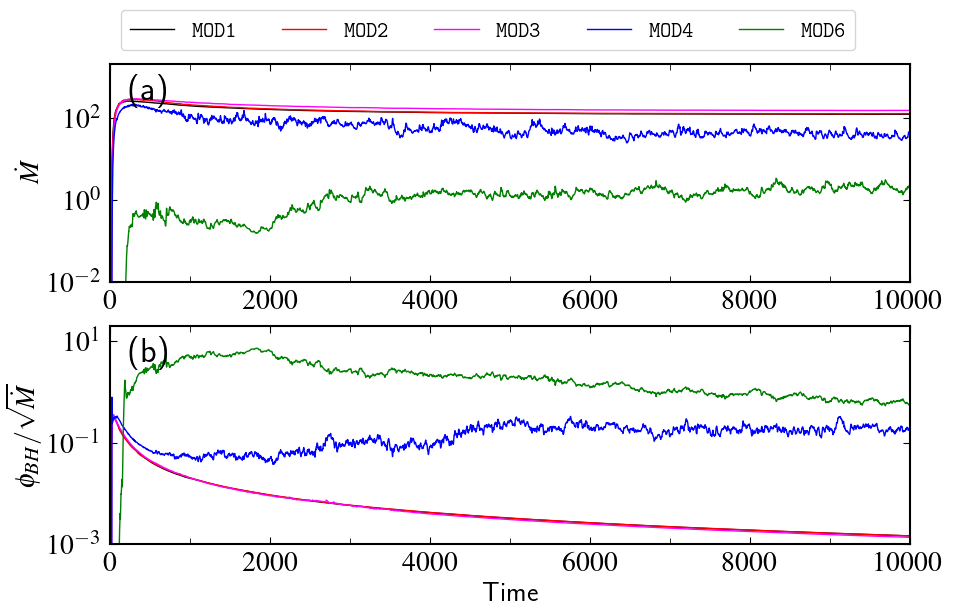}
    \caption{Time evolution of the accretion rate ($\dot{M}$, panel (a)) and the normalized magnetic flux ($\phi_{\rm BH}/\sqrt{\dot{M}}$, panel (b)) calculated at the horizon for different models with different angular momentum. See text for more details.}
    \label{fig:acc-mag}
\end{figure}
\begin{figure}
    \centering
    \includegraphics[width=1.0\linewidth]{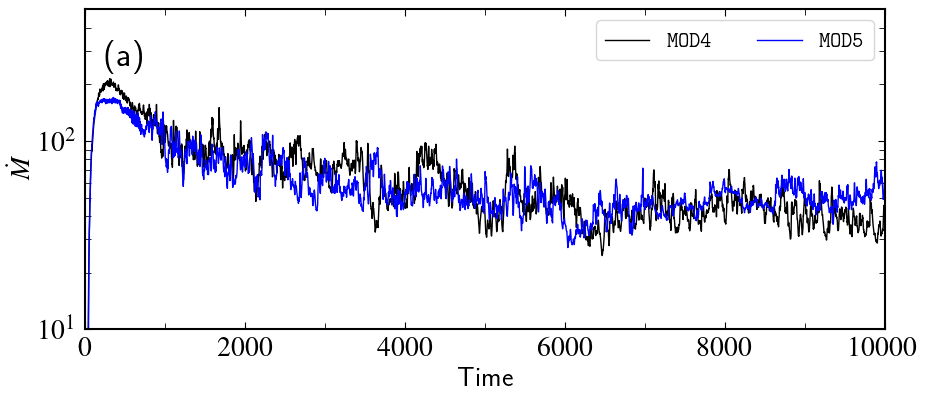}
    \caption{Time evolution of the accretion rate ($\dot{M}$) model \texttt{MOD4} (with magnetic field) and \texttt{MOD6} (without magnetic field). See text for more details.}
    \label{fig:comp-acc}
\end{figure}

In this section, we investigate the impact of low angular momentum flows on the time-series properties. To do that, we plot the accretion rate ($\dot{M}$) and the normalized magnetic flux ($\phi_{\rm BH}/\sqrt{\dot{M}}$) calculated at the horizon (following \cite{Porth-etal2017}) for different models in figure~\ref{fig:acc-mag} (panels a and b, respectively). In the figure, black, red, magenta, blue, and green lines correspond to \texttt{MOD1-4} and \texttt{MOD6}, respectively. They are calculated for ${\cal F}=0.2, 0.3, 0.4, 0.5$, and $1.0$, respectively. 

In panel (a) of Fig.~\ref{fig:acc-mag}, we observe distinct changes in the feature of the accretion rate profile with the decrease of angular momentum. For the case of \texttt{MOD6} (magenta), the angular momentum is $193.4\%$ of $\lambda_{\rm ms}$ and $\lambda_0>\lambda_{\rm mb}$, also the flow is gravitationally bound (${\cal E}=-hu_t<1$, see Appendix A for detail). Therefore, flow cannot cross the marginally stable orbit unless it loses some angular momentum or gain energy via turbulence in the torus driven by magneto-rotational instability (MRI) \citep[][see section \ref{sec:result2}.1 for more detail]{Balbus-Hawley1991,Balbus-Hawley1998}. Note that for this model, the minimum angular momentum at the inner edge ($\lambda_{\rm min}\sim3.44$) is also greater than both $\lambda_{\rm ms}$ as well $\lambda_{\rm mb}$.
As a result, we observe accretion across the horizon after simulation time $t>250\,t_g$ and the value gradually increases and reaches a quasi-steady state. However, for the other models, the angular momentum is lower than that of the marginally stable angular momentum. Therefore, the flow directly plunges into the horizon, and we see a very high value of the accretion rate immediately.
For model \texttt{MOD4} (blue), the angular momentum is slightly lower than that of the $\lambda_{\rm ms}$. 
As a result, the pressure (thermal and magnetic) developed close to the black hole could still give us an accretion rate profile qualitatively similar to model \texttt{MOD6} with variability. Nonetheless, it is clear from the plots that they are quantitatively very different.
On the contrary, other models have very low angular momentum as compared to $\lambda_{\rm ms}$, and therefore, we see a very smooth accretion rate profile without any variability signature. To study it quantitatively, we calculate the variability magnitude, $\delta\dot{m}/\langle\dot{m}\rangle=({\rm max}(\dot{m})-{\rm min}(\dot{m}))_{\delta t}/\langle\dot{m}\rangle$, with $\dot{m}=\log_{10}\dot{M}$, for models \texttt{MOD1}, \texttt{MOD2}, \texttt{MOD3}, \texttt{MOD4}, and \texttt{MOD6} within simulation time $t=8000-10000\,t_g$.
They are obtained as $0.24\%$, $0.25\%$, $0.24\%$, $24.0\%$, and $168.0\%$, respectively. In terms of the magnitude of the accretion rate, low angular momentum flow can fall onto the black hole faster due to a weaker centrifugal barrier. Accordingly, we observe higher values of the accretion rate for lower angular momentum cases, and they have quite similar profiles throughout the simulation time.  

Similar to the accretion rate, we see distinct features in the normalized magnetic flux profiles at different angular momentum limits in panel Fig.~\ref{fig:acc-mag}b. For high angular momentum flow (model \texttt{MOD6}), the magnetic flux grows with simulation time. However, due to limited resolution, the magnetic flux starts to drop after simulation $t>2000\,t_g$. For model \texttt{MOD4}, initially we observe a decrease in the magnetic flux. However, as time passes, it increases again and reaches a steady value. On the contrary, for the cases with lower-angular momentum (\texttt{MOD1-3}), the magnetic flux decreases monotonically with time, and the profiles look quite similar to them. Overall, due to the longer in-fall time than MRI growth time, higher angular momentum cases can accumulate more magnetic flux near the event horizon. Finally, we can conclude that all our models are in the regime of standard and normal evolution (SANE); they do not have enough magnetic flux to become magnetically arrested disk (MAD) regime in this simulation time \citep [e.g.,][]{Tchekhovskoy-etal2010,Porth-etal2021,Mizuno-etal2021}.

Next, we compare the time-series properties of the flow with and without magnetic fields. To do that, in Fig.~\ref{fig:comp-acc}, we present the accretion rates for models \texttt{MOD4} (with magnetic fields) and \texttt{MOD5} (without magnetic fields). The accretion rate profiles seem qualitatively similar; however, the flow properties may still be different, which we will study in the next section. Note that, for model \texttt{MOD4}, the magnetic flux around the event horizon is only about $\phi_{\rm BH}/\sqrt{\dot{M}}\sim0.2$ at the end of the simulation. This implies that the magnetic field for the model is very weak, and accordingly, we expect similar results for both models. We will study the impacts of strong magnetic fields on low angular momentum flows in our upcoming studies. 

The time-series analysis suggests that the flow properties for \texttt{MOD1}, \texttt{MOD2}, and \texttt{MOD3} are quite similar. Accordingly, for the sake of better explanations in the later part of the paper, we call them very low angular momentum cases and consider \texttt{MOD1} as a representative case for them. Further, we call \texttt{MOD4} and \texttt{MOD6} intermediate-angular momentum and high-angular momentum cases, respectively.

\section{Flow properties}\label{sec:result2}
\begin{figure*}
    \centering
    \includegraphics[scale=0.35]{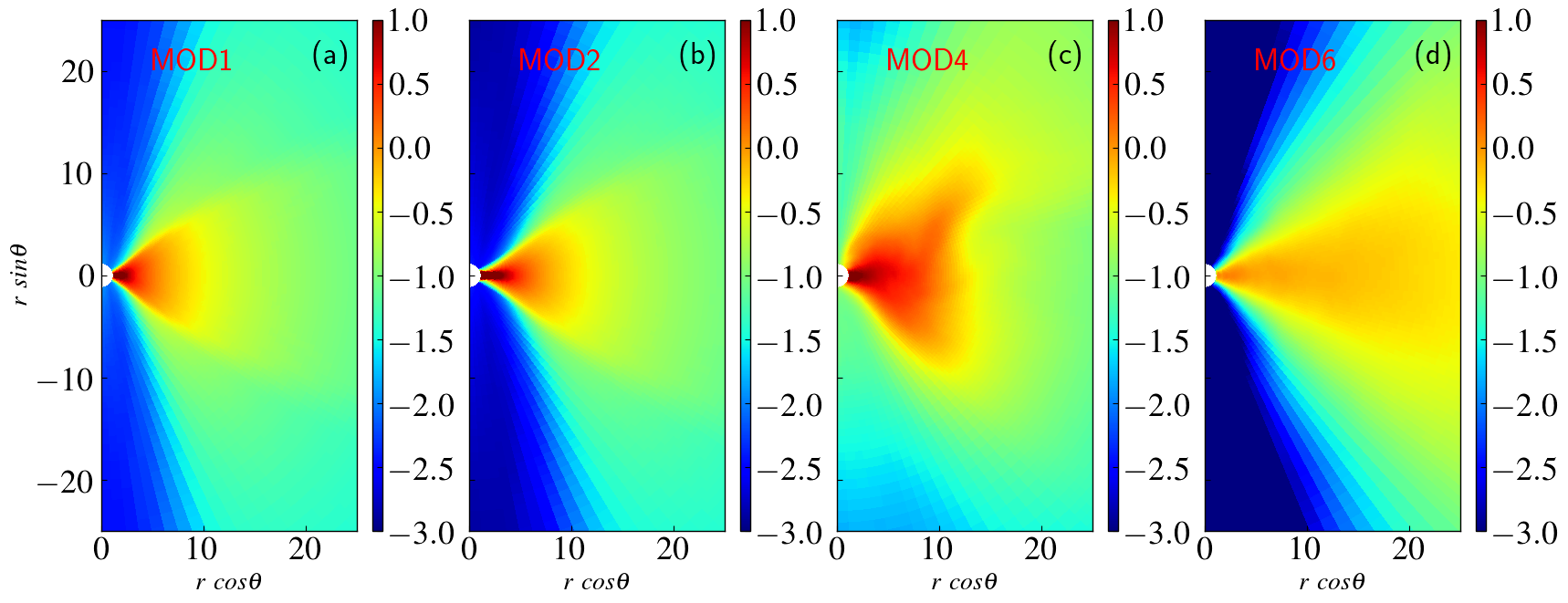}
    \caption{Time average density distribution for a slice at $\phi=0^\circ$ within simulation time $t=8000-10000\,t_g$ for different models. See text for more details.}
    \label{fig:density-3d}
\end{figure*}

In this section, we investigate the detailed flow properties of low angular momentum flows. To do that, in Fig.~\ref{fig:density-3d}, we present the time average density distribution for a slice at $\phi=0^\circ$ within simulation time $t=8000-10000\,t_g$ for different models. Panels (a)-(d) correspond to models \texttt{MOD1, MOD2, MOD4,} and \texttt{MOD6}, respectively. Remember that these models are arranged in an increasing trend of the magnitude of angular momentum (see table \ref{tab-01}). The panels show clear differences in the density distribution while increasing angular momentum. For very low angular momentum (\texttt{MOD1}, panel (a)), we observe a high-density conical region close to the black hole. 
The conical shape of the disk essentially indicates a vertically pressure-supported structure, where the pressure gradient balances the gravitational attraction normal to the disk. In such a case, the disk height can be calculated as $H\propto \sqrt{p_g/\rho}\times r^{3/2}$ with $r\gg1$ \citep{Shakura-Sunyaev1973,Frank-etal2002}. In the case of temperature, $p_g/\rho=\Theta$ is proportional to $r^{-1}$, disk height becomes $H \propto r$.
In later sections, we demonstrate that for very low angular momentum cases, $\Theta\propto r^{-1}$ (see section \ref{sec:result3} for detail).
With an increase in angular momentum, the opening angle of the cone increases (\texttt{MOD2}, panel (b)). However, when angular momentum is comparable to marginally stable angular momentum (\texttt{MOD4}, panel (c)), we do not observe any conical shape. Instead, we see a high-density clump of matter around the black hole, indicating the presence of turbulent flows. On the contrary, for very high angular momentum (\texttt{MOD6}, panel (d)), the density distribution is much smoother than \texttt{MOD4}. However, we do not observe a clear conical shape for \texttt{MOD6} as seen in very low angular momentum models (\texttt{MOD1, MOD2}).

\begin{figure*}
    \centering
    \includegraphics[scale=0.35]{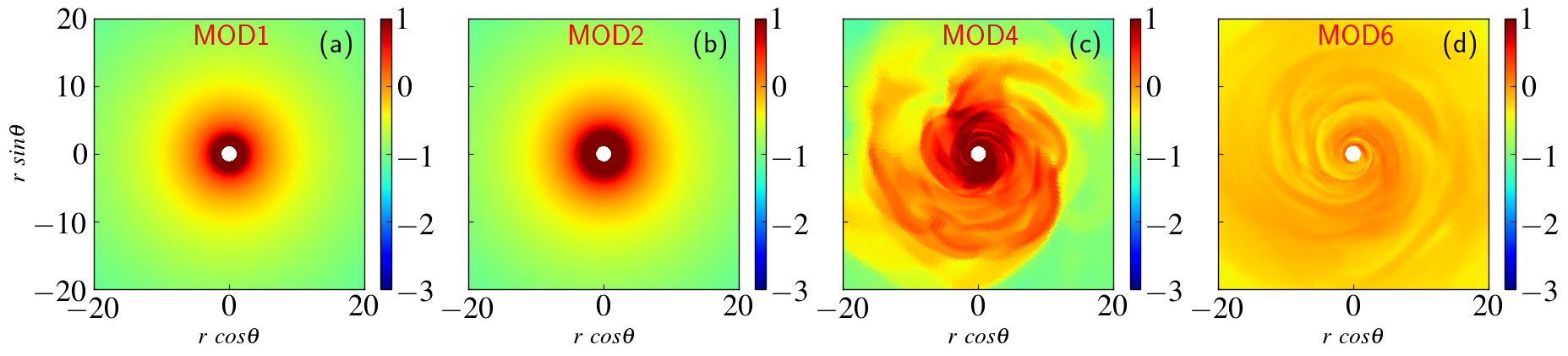}
    \caption{Density distribution of equatorial plane at simulation time $t=10000\,t_g$ for different models. See text for more details.}
    \label{fig:density-2d}
\end{figure*}

The difference in the density distribution can be comprehended by observing it on the equatorial plane for different simulation models, which we show in Fig.~\ref{fig:density-2d}. In this figure, we plot the density distribution at simulation time $t=10000\,t_g$. The panels are distributed exactly in the same fashion as in Fig.~\ref{fig:density-3d}. We note that, with the increase in angular momentum, the radius of the high-density compact region close to the black hole increases, which suggests competition between gravitational pull and centrifugal force. For \texttt{MOD4} (panel c), we see the spiral density structures in the accretion flow. Earlier low angular momentum GRHD simulations also reported similar spiral structures or density jumps associated with local shock transitions \citep{Olivares-etal2023}. This indicates that our simple adaptation of FM torus could seemly reproduce earlier results from a much more complex setup. These kinds of structures are usually observed in the near-horizon region of MAD flow \citep[e.g.,] []{Dexter-etal2020,Porth-etal2021,Scepi-etal2022,Ripperda-etal2022}. They are useful to understand flaring events in the light curve of Sgr~A$^*$ \citep{Dexter-etal2020,Scepi-etal2022,Ripperda-etal2022}; however, the timing properties cannot be reproduced properly with MAD models. However, in this case, we observe spiral structures slightly far from the horizon. This may help in understanding the timing properties of Sgr~A$^*$ consistently. Note that this flow is not in the MAD regime. The magnetic flux around the horizon is about $\phi_{\rm BH}/\sqrt{\dot{M}}\sim 0.2$. Therefore, we do not have a strong jet activity, which is also expected from the Sgr~A$^*$ \citep{Royster-etal2019,Yusef-Zadeh-etal2020}. To confirm these claims, we need explicit radiation calculations by general-relativistic radiation transfer codes, which we plan to do in subsequent studies. Finally, for the usual high-angular momentum flows (\texttt{MOD6}, panel d), we see a more seminal flow structure than that in low angular momentum flows. 

\begin{figure*}
    \centering
    \includegraphics[scale=0.35]{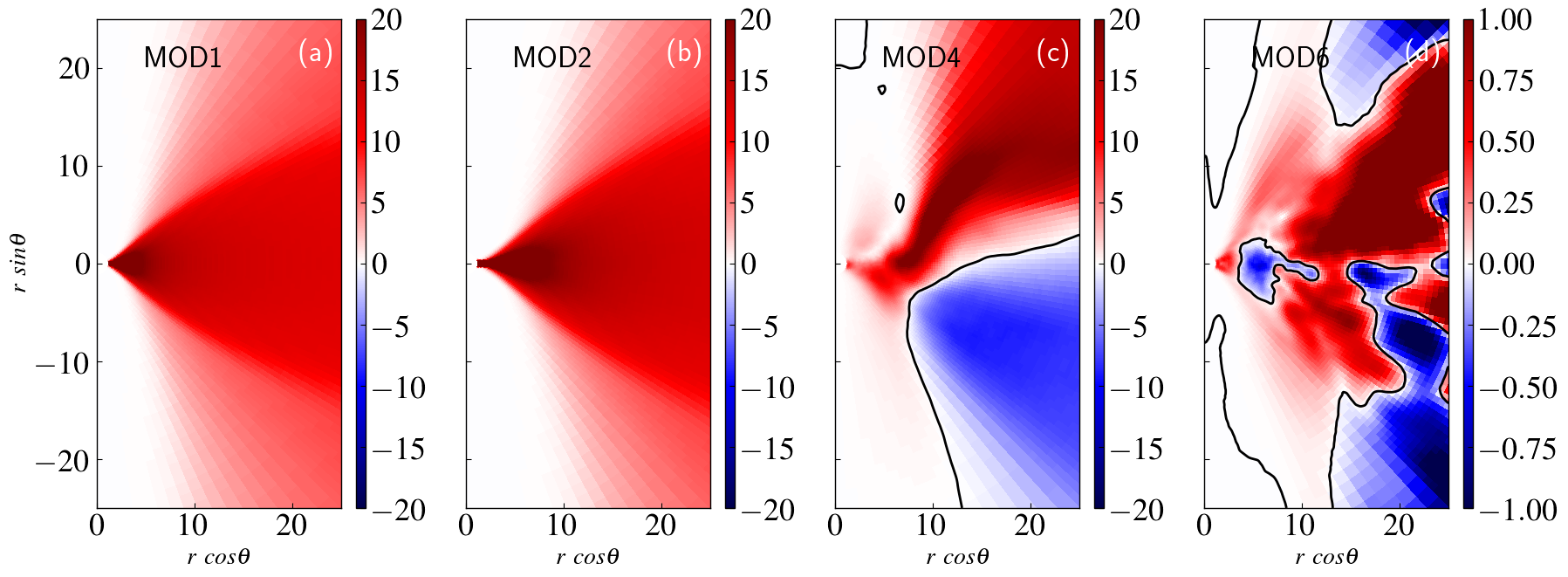}
    \caption{Time average mass flux $(\sqrt{-g}\rho u^r)$ distribution for a slice at $\phi=0^\circ$ within simulation time $t=8000-10000t_g$ for different models. Black solid line corresponds to $\sqrt{-g}\rho u^r=0$.  See text for more detail.}
    \label{fig:mdot-3d}
\end{figure*}

\begin{figure*}
    \centering
    \includegraphics[scale=0.35]{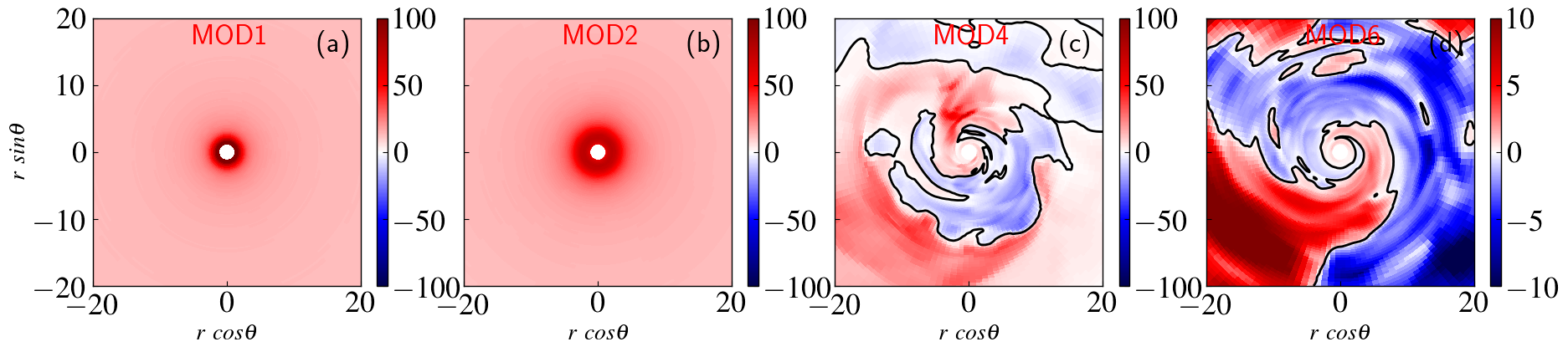}
    \caption{Mass flux $(\sqrt{-g}\rho u^r)$ distribution at equatorial plan at simulation time $t=10000t_g$ for different models. Black solid line corresponds to $\sqrt{-g}\rho u^r=0$. See text for more detail.}
    \label{fig:mdot-2d}
\end{figure*}
\begin{figure*}
    \centering
    \includegraphics[scale=0.35]{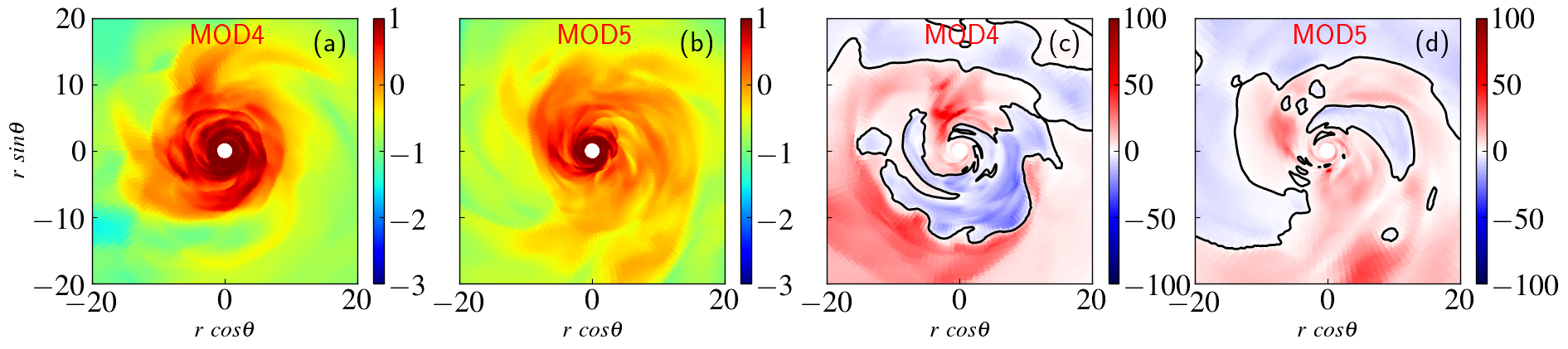}
    \caption{Comparison of density distribution (a,b) and mass flux distribution (c,d) for magnetized (\texttt{MOD4}) and unmagnetized (\texttt{MOD5}) on the equatorial plane at time $t=10000t_g$.}
    \label{fig:den_mdot_comp}
\end{figure*}
Next, we investigate the mass flux distribution to study inflow/outflow activities with the increase in angular momentum.  Fig.~\ref{fig:mdot-3d} shows the time average mass flux $(\sqrt{-g}\rho u^r)$ distribution for a slice at $\phi=0^\circ$ within simulation time $t=8000-10000\,t_g$ for different models. In the panels, the black solid line corresponds to $\sqrt{-g}\rho u^r=0$. In panels (a) and (b), we do not observe any blueish regions, indicating no outflow in the very low angular momentum ranges. Similar to the density distributions, we see a distinct conical structure for mass flux as well. However, for \texttt{MOD4} (panel c), we observe inflow as well as outflow regions, showing the transport of angular momentum. Interestingly, we do not observe bipolar outflows. On the contrary, the outflow region is toward the equatorial region. Moreover, inflow is also tilted with respect to the equatorial plane. This suggests the formation of a tilted disk close to the black hole due to the shearing instabilities in the inflow-outflow boundary. We will discuss more detail about it subsequently. 
Finally, for the usual high-angular model, the angular momentum needs to be transported from all the regions through MRI for accretion to take place. Accordingly, we observe the bluish region in a more or less symmetric manner. We also observe bipolar outflows and jets around the polar axis. Similarly, in Fig.~\ref{fig:mdot-2d}, we show the mass flux on the equatorial plane analogously to Fig.~\ref{fig:mdot-3d}. The panels in the figure also suggest that the very low angular momentum flow is seminal in nature. With the increase in angular momentum, we observe inflow-outflow boundaries on the equatorial plane. This feature is quite similar for \texttt{MOD4} and \texttt{MOD6}.

At the end, we study the differences between magnetized and unmagnetized flow. To do that, we show the density (a,b) and mass flux (c,d) distribution at the equatorial plane in Fig.~\ref{fig:den_mdot_comp}, where models \texttt{MOD4} and \texttt{MOD5} correspond to magnetized and unmagnetized flows, respectively. Qualitatively, we see quite similar features in magnetized and unmagnetized flows. This is due to the presence of a very weak magnetic field. However, in the density distribution, we observe that the spiral filaments are more prominent but compact for magnetized flows than those of the unmagnetized flows. This is because, due to magnetized flow, angular momentum transport is efficient. Therefore, the inner part has lower angular momentum for magnetized flows than that of unmagnetized flows. As a result, the mass flux distributions in magnetized flows show slightly higher values than those of the unmagnetized flow (see dark red and blue colors in panel c). These differences may be more drastic with the increase in magnetic field strength. In the next section, we discuss in more detail differences in the flow properties in terms of vertically integrated quantities. 

\subsection{Development of turbulent structure}
\begin{figure}
    \centering
    \includegraphics[scale=0.35]{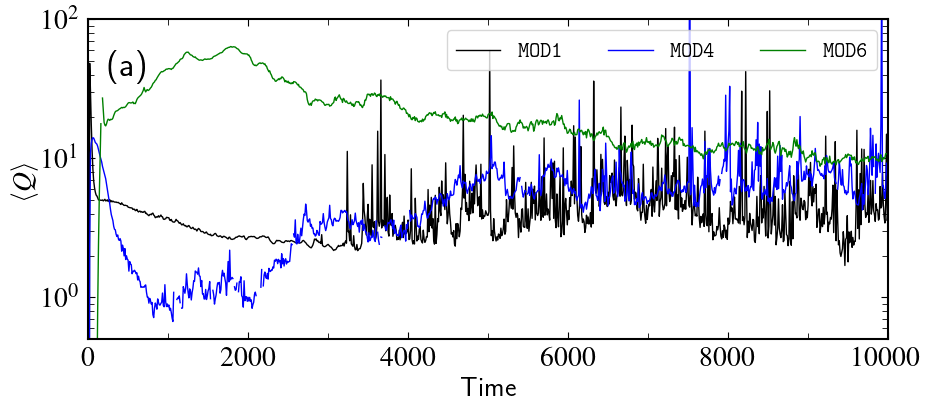}
    \caption{Time evolution of average MRI quality factor $\langle Q \rangle$ for different simulation models with different angular momentum.}
    \label{fig:Qtime}
\end{figure}

In all the simulation models, we observe that the flow becomes turbulent with the increase in angular momentum. In this section, we want to understand the detailed process of its formation and how flow develops spiral filaments in the accretion flow. We note that the turbulent structure is present in magnetized as well as unmagnetized flow. Therefore, we expect the source of it to have a non-magnetic origin. However, MRI could still influence the turbulence in the flow. In Fig.~\ref{fig:Qtime}, we show the time evolution of grid-averaged MRI quality factor for the fastest growing modes: $\langle Q \rangle=(1/3)(\langle Q_r \rangle + \langle Q_\theta \rangle + \langle Q_\phi \rangle)$ (see \cite{Takahashi2008,Porth-etal2019} for detail calculation), where averaging is performed within $r<100\,r_g$ nearby the equatorial plane within $\pm \pi/5$. The figure suggests that depending on the angular momentum, the fastest-growing MRI wavelength grows differently. For the highest angular momentum, the wavelength is longer and can be resolved with a higher value of $\langle Q \rangle$. However, with a decrease in angular momentum, the infall timescale becomes shorter than the MRI growth timescale. 
As a result of lower angular momentum, the MRI quality factor becomes low.
Note that to resolve the MRI adequately, at least $\langle Q \rangle \gtrsim 6$ is needed \citep{Sano-etal2004}. Except in model \texttt{MOD6}, the accretion process can naturally happen as the angular momentum is lower than that of the marginally stable angular momentum and transits to a quasi-steady/steady state in later simulation time. In this case (\texttt{MOD6}), the MRI quality factor becomes $\sim 10$ at the end of the simulation, where accretion is driven by resolved MRI. Therefore, the qualitative results from our simulation models are not expected to alter with the increase in resolution. 
Since, with resolution, the magnetic pressure will grow due to well-resolved MRI, we may expect quantitatively similar results at slightly lower angular momentum as compared to the current values mentioned in this work. Moreover, for high resolution, the numerical dissipation will be reduced. Along with the extra magnetic pressure at high resolution, we expect to see a reduction in the overall accretion rate. We plan to do such simulations in the future and report elsewhere.

\begin{figure}
    \centering
    \includegraphics[scale=0.35]{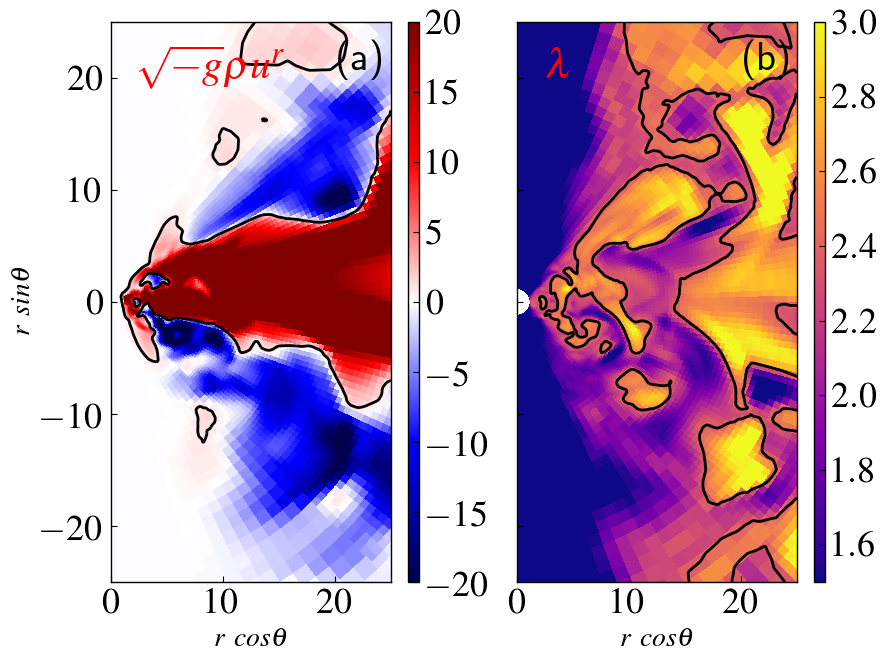}
    \caption{The distribution of mass flux $(\sqrt{-g}\rho u^r)$ and specific angular momentum $(\lambda=-u_\phi/u_t)$ at simulation time $t=1500t_g$ and $\phi=0$ for model \texttt{MOD4}. Contours in panel (a) and (b) correspond to $u^r=0$ and $\lambda=\lambda_{\rm ms}$. See text for more details.}
    \label{fig:turb}
\end{figure}

Finally, to understand the origin of the turbulent nature of the accretion flow in models \texttt{MOD4} and \texttt{MOD5}, we show the distribution of mass flux $(\sqrt{-g}\rho u^r)$ and specific angular momentum $(\lambda=-u_\phi/u_t)$ in panels (a) and (b) of Fig.~\ref{fig:turb}, respectively, during the initial phase of the simulation at $t=1500\,t_g$ for model~\texttt{MOD4}. For the intermediate angular momentum case, gravity and centrifugal force are more or less equal at the ISCO. Accordingly, the increase in pressure close to the black hole could push some outflow in a bipolar direction for an intermediate range of angular momentum. As a result, we observe the development of shear between inflow and outflow, which we can clearly see in panel (a) of Fig.~\ref{fig:turb}. The black solid line ($u^r=0$) in Fig.~\ref{fig:turb}a indicates the inflow-outflow boundary. Velocity shear at the boundary triggers the excitation of instability and turbulence. Due to the turbulence developed by the shear instability, angular momentum is transported outward. That further enhances the outflow close to the black hole. The consequences can be seen in the angular momentum distribution. We observe islands of high-angular momentum in the outflow region in panel (b) of Fig.~\ref{fig:turb}. In Fig.~\ref{fig:turb}b, the solid lines show the contour of $\lambda=\lambda_{\rm ms}$. This outflow does not have enough kinetic energy to leave the system (i.e., is bounded). Thus, it falls back on the accretion flow far from the black hole. This recurrent process destabilized seminal flow near the equatorial plane, and finally, we see a spiral filament structure in the accretion flow. In the very low angular momentum cases (\texttt{MOD1}-\texttt{MOD3}), the centrifugal force is always much smaller than that of gravity. Accordingly, pressure is not enough to push outflow, and shear instability is never triggered. As a result, we do not observe any spiral structure. Nonetheless, there is a possibility of having such a structure even in low angular momentum flow with an increase in magnetic pressure. That we plan to study in the future.

\section{Vertically averaged structure}\label{sec:result3}

\begin{figure}
    \centering
    \includegraphics[scale=0.7]{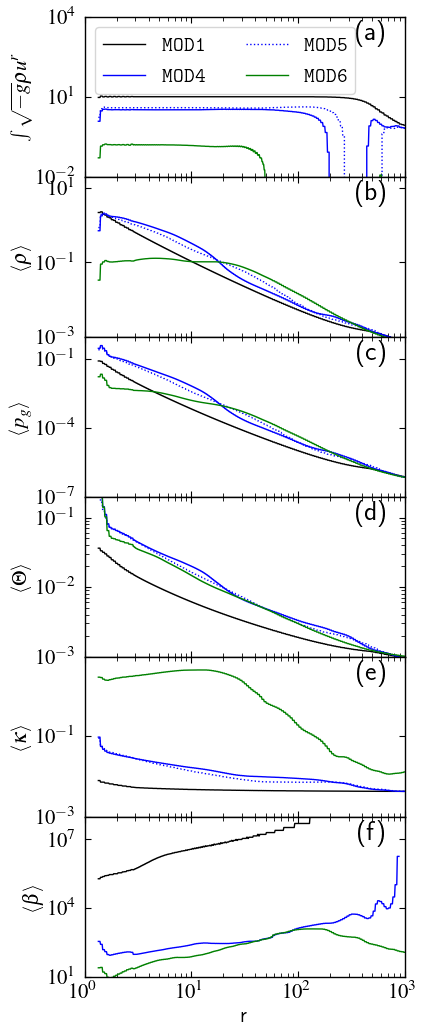}
    \caption{Radial distribution of vertically integrated mass flux ($\int \sqrt{-g} \rho u^r$), and averaged density ($\langle \rho \rangle$), gas pressure ($\langle P \rangle$), temperature ($\langle \Theta=P/\rho \rangle$), specific entropy ($\langle \kappa \rangle$), and plasma-$\beta$ ($\langle \beta \rangle$) for different simulation models. See text for details.}
    \label{fig:int-flow-prop}
\end{figure}

In this section, we study the influence of angular momentum on the time-averaged vertically integrated structure of the accretion flow close to the black hole. To do that, we calculate the average quantities $\langle X \rangle$ following:
\begin{align}
    \langle X \rangle = \frac{\int \sqrt{-g} X  d\theta d\phi}{\int \sqrt{-g} d\theta d\phi},
    \label{eq:avg}
\end{align}
where $X$ is a time-averaged quantity between $t=8000-10000\,t_g$. Following Eq.~\ref{eq:avg}, we plot the averaged value of density ($\langle \rho \rangle$), gas pressure ($\langle p_g \rangle$), flow temperature ($\langle \Theta=p_g/\rho \rangle$), specific entropy ($\langle \kappa \rangle$), and plasma-$\beta$ ($\langle \beta \rangle$) as a function radius in panels (b)-(f) of Fig.~\ref{fig:int-flow-prop}, respectively. In panel (a) of Fig.~\ref{fig:int-flow-prop}, we show vertically integrated $\int \sqrt{-g}\rho u^r$ (without vertical averaging) to study the inflow/outflow equilibrium of the models. In the panels, black, blue, and green correspond to \texttt{MOD1} (very low angular momentum), \texttt{MOD4} (moderate angular momentum), and \texttt{MOD6} (high angular momentum), respectively. Additionally, we also plot the same quantities for the unmagnetized model (\texttt{MOD5}) by dotted blue lines for comparison.

In panel (a) of Fig.~\ref{fig:int-flow-prop}, we show an inflow/outflow equilibrium diagram for different angular momentum limits. In the lowest angular momentum case, the value of the equilibrium of mass flux is the highest (\texttt{MOD1}). Subsequently, in the higher angular momentum case, it becomes lower. This suggests that low angular momentum flows can reach the event horizon without any barrier. We also note that the equilibrium reaches up to $r\sim400\,r_g$, $\sim 100\,r_g$, and $\sim2\,0r_g$ for models \texttt{MOD1, MOD4}, and \texttt{MOD6}, respectively, although the simulation time is the same, i.e., $t=8000-10000\,t_g$. Accordingly, for the following discussions, we will only compare results within $r<20\,r_g$ for consistency.

In panel (b) of Fig.~\ref{fig:int-flow-prop}, we observe that the nature of the density profile changes significantly with the increase of angular momentum. Interestingly, the intermediate range of angular momentum has the highest density with a radius of $r<20\,r_g$. Additionally, high-angular momentum flow has the lowest density close to the black hole. To quantify the differences, we fit the density with a power-law profile $\rho\propto r^{\alpha_\rho}$. We find that the values of power-law index $\alpha_\rho$ are $\alpha_\rho=-1.54$ ($\sim-3/2$), $-1.00$, and $-0.06$ ($\sim0$), for models \texttt{MOD1}, \texttt{MOD4}, and \texttt{MOD6}, respectively. Similar to panel Fig.~\ref{fig:int-flow-prop}b, the averaged pressure also shows similar behavior in Fig.~\ref{fig:int-flow-prop}c. We observe maximum pressure in the intermediate range of angular momentum. The power-law index that could fit $p_g\propto r^{\alpha_p}$ is obtained as $\alpha_p=-2.49$ ($\sim -5/2$), $-1.74$, and $-0.71$ for models \texttt{MOD1, MOD4}, and \texttt{MOD6}, respectively. Similar to pressure and density, the temperature profiles also show maximum values for the intermediate range of angular momentum. However, for the temperature profile, the nature of the plots is quite similar. Although we see some changes in the power law $\Theta\propto r^{\alpha_\Theta}$, where $\alpha_\Theta=-1$, $-0.75$, and $-0.75$ for very low, intermediate, and high-angular momentum cases, respectively.

Subsequently, in panels (e) and (f) of Fig.~\ref{fig:int-flow-prop}, we show entropy ($\langle \kappa \rangle$) and plasma-$\beta$ ($\langle \beta \rangle$) profiles for different limits of angular momentum. Panel Fig.~\ref{fig:int-flow-prop}e shows that higher-angular momentum has a higher entropy of the flow. The solutions are also not isentropic; the entropy increases as flows proceed towards the black hole. In the very low angular momentum case, the entropy throughout is more or less constant. Therefore, we expect the properties of flows for very low angular momentum cases to be similar to isentropic semi-analytic solutions \citep{Bondi1952,Chakrabarti1989}. Similarly, the strength of the magnetic field is higher for higher angular momentum flows. Since the power-law indices for density and pressure are different at different angular momentum limits, the power-law for entropy is also different. We do not calculate them explicitly here. However, it is interesting to check the power-law indices for plasma-$\beta$ ($\beta\propto r^{\alpha_\beta}$). We find that $\alpha_{\beta}=1.66$ ($\sim 5/3$), $0.54$ ($\sim1/2$) and $1.31$ ($\sim 4/3$) for models \texttt{MOD1, MOD4}, and \texttt{MOD6}, respectively.

Earlier, such power-law indices were often used to model accretion flows around a black hole using self-similarity. The power-law indices for the very low angular momentum case show extraordinary similarities to the values proposed by \cite{Narayan-Yi1995} for ADAF solutions. Subsequently, \cite{Blandford-Begelman2004} also calculated the power-law indices, considering a more general scenario involving outflows. They proposed that the density and pressure must satisfy $\rho\propto r^{n-3/2}$ and $p_g\propto r^{n-5/2}$ for self-similarity in the axisymmetric assumption, where $n$ depends on the outflows. By comparing the power-law indices, we find that $n=0$ matches exactly for the very low angular momentum case. $n=0$ is also reported for low angular momentum, magnetized, and transonic accretion flows \citep{Mitra-etal2022}. Recently, \cite{Olivares-etal2023} have performed similar calculations for transonic relativistic hydrodynamic simulations. They reported density variations $\rho \propto r^{-3/2}$. Note that here we have only shown plots for \texttt{MOD1}; nonetheless, similar to this model, \texttt{MOD2} and \texttt{MOD3} also follow $n\sim0$ for both density and pressure. This is reasonable because these models do not have any outflow, and flow properties are also axisymmetric (see panels (a) and (b) of Fig.~\ref{fig:density-3d}-\ref{fig:mdot-2d}). These conditions do not hold for intermediate as well as high-angular momentum flows. Accordingly, such self-similarity does not hold. With such self-similarity, we expect $\alpha_\Theta=-1$, and we observe its deviation with the increase of angular momentum. A similar trend is also reported in \cite{Olivares-etal2023}. It has been shown that for MAD models, density varies as $\rho\propto r^{-1}$, which, interestingly, is exactly the same for the intermediate angular momentum flow model, although it is in the SANE regime. The wind-fed hydrodynamic simulations done by \cite{Ressler-etal2018} show density variations $\rho \propto r^{-1}$, which is exactly the same as our intermediate angular-momentum case.

Finally, comparing the solid and dotted lines in all the panels, we investigate the flow properties between magnetized and unmagnetized flows. We have already mentioned that, due to the weak nature of the magnetic field, the difference in all the profiles is minimal. Although we observed slightly lower values for density, pressure, temperature, and entropy for unmagnetized flow, we expect quite similar power-law indices for magnetized and unmagnetized flows. At the same time, it will be interesting to study it by gradually increasing the magnetic field strength. We plan to do such a study in the future.

\begin{figure}
    \centering
    \includegraphics[scale=0.55]{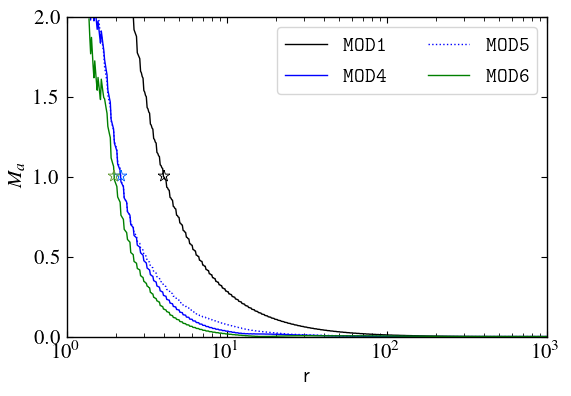}
    \caption{Radial distribution of Mach number $(M_a=\Tilde{v}_{rad}/a_s)$ at the equatorial plane for different simulation models having different angular momentum. See text for more details.}
    \label{fig:mach-number}
\end{figure}

\subsection{Sonic state of the models}

Finally, we study the Mach number variation with different angular momentum, which gives the sonic state of the flow. Here, we calculate radial velocity $\Tilde{v}_{rad}$ in corotating frame in the Boyer-Lindquist coordinate following $\Tilde{v}_{rad}^2=\gamma_\phi^2u^ru_r/(-u_tu^t)$, where $\gamma_\phi^2$ is the Lorentz factor for rotation, i.e., $\gamma_\phi^2=1/(1+u_\phi u^\phi/u_tu^t)$ \citep[e.g.,][]{Dihingia-etal2018a}. Additionally, we calculate the sound speed as $a_s^2 = \Gamma p_g/(e + p_g)$, where $e$ and $\Gamma=4/3$ are internal energy and the adiabatic index of the flow, respectively.
Fig.~\ref{fig:mach-number} shows the Mach number ($M_a$) on the equatorial plane for models \texttt{MOD1} (very low angular momentum, black solid line), \texttt{MOD4} (intermediate angular momentum, blue solid line), \texttt{MOD5} (unmagnetized intermediate angular momentum, dotted blue line), and \texttt{MOD6} (high angular momentum, solid green line). The stars on the curve correspond to the sonic point (sonic surface in 3D visualization) for the flows, where flows make a transition from subsonic ($M_a<1$) to supersonic ($M_a>1$). The plots suggest that with the decrease in angular momentum, the location of the sonic point moves far from the black hole. The location of sonic points for \texttt{MOD1, MOD4}, and \texttt{MOD6} are obtained as $r_c=4.06\,r_g$, $2.16\,r_g$, and $1.97\,r_g$, respectively. Note that the innermost stable circular orbit (ISCO) for this Kerr black hole is $r_{\rm ISCO}=2.04\,r_g$. Thus, with high angular momentum, a sonic point resides inside the ISCO radius, whereas with a decrease in angular momentum, a sonic point may form outside the ISCO radius. 
For unmagnetized flow, the difference in the Mach number profile is minimal. These kinds of solutions contain only a single sonic point; however, it is possible to have two sonic points in a single solution if there is a standing shock in the solution \cite[e.g.,][]{Fukue1987,Chakrabarti1989}.  The ample number of semi-analytical studies have shown the presence of such solutions with standing shocks in different scenarios \citep[references therein]{Das2007,Chakrabarti-etal2015,Dihingia-etal2018,Dihingia-etal2019a,Dihingia-etal2019b,Dihingia-etal2020}. To date, such standing shock solutions cannot be detected with the GRMHD simulation framework. However, transient shocks could be possible from an axisymmetric truncated magnetized thin accretion disk \citep{Dihingia-etal2022}. Due to differences in the physical nature of the flow considered for current work, we do not observe any such transient shocks in this study. However, this preliminary study hints that, for some range of parameters, standing or oscillatory shocks could be possible. Earlier parameter space surveys for ${\cal E}(=-hu_t)$ and ${\cal L}(=hu_\phi)$ (or specific angular momentum $\lambda(=-u_t/u_\phi)$) suggest that shock is possible only within a restricted range of these parameters (e.g., \cite{Chakrabarti1996,Kumar-Chattopadhyay2017,Dihingia-etal2019a,Dihingia-etal2019b}). In semi-analytic models, these ranges are calculated considering a thin disk. However, in our cases, the flow is not geometrically thin. To get a global standing/oscillating shock, we need to perform our simulation considering suitable values of these parameters. In this study, we have not explored different ranges of initial values of ${\cal E}$ yet. Since our goals for this work do not consider it, nonetheless, it is important to carry it out in the future. Accordingly, we plan to do more parameter space surveys in our upcoming studies to investigate whether shock solutions are possible in 3D GRMHD simulations or not.

\section{Summary and discussion}\label{sec:summary}

In this study, we focused on investigating the flow properties at different limits of angular momentum. Primarily, we considered very low angular momentum, intermediate angular momentum, and high angular momentum cases and performed 3D GRMHD simulations around the Kerr black hole with moderate resolution. We employ the initial conditions of the FM torus to perform our simulations. Interestingly, we could reproduce results from a much more complicated simulation setup for low angular momentum accretion flows, e.g., \cite{Ressler-etal2018,Olivares-etal2023}. Here, we list our major findings in point-wise order:
\begin{itemize}
    \item We found that lower-angular momentum accretion flow has a higher saturated accretion rate. Higher angular momentum flow has a higher variability magnitude $(\delta\dot{m}/\langle \dot{m}\rangle)$. For the variability of very low angular momentum flows, the magnitude tends to zero, or the accretion rate profiles become absolutely smooth. Similarly, the normalized magnetic flux ($\phi/\sqrt{\dot{M}}$) does not saturate for very low angular momentum cases. For the intermediate angular momentum case, normalized magnetic flux saturates to lower values as compared to the high-angular momentum case.
    
    \item By lowering angular momentum to an intermediate value of $\sim97\%\lambda_{\rm ms}$, we observe accretion flow with distinct spiral filaments. It is absent if angular momentum is very low or very high. The high-density region close to the black hole becomes compact with a decrease in angular momentum.
    
    \item We found that angular momentum impacts the onset of outflows/jets from the accretion flow. In the intermediate angular momentum range, we observe outflow more towards the equatorial plane. However, for very low angular momentum cases, outflow/jet also ceases to exist. In such cases, we observe pure inflows.

    \item The flow properties of different angular momentum fit with different power-law indexes. In general, we can fit density and pressure with $\rho\propto r^{n-3/2}$ and $p_g\propto r^{n-5/2}$ \citep{Blandford-Begelman2004} for very low angular momentum axisymmetric cases without any outflow with $n=0$. However, with the increase in angular momentum, the accretion develops non-axisymmetric flows. Therefore, such simple self-similarity does not hold for intermediate as well as high-angular momentum cases. 
    
    \item In our study, we also investigated the sonic properties of the simulation models. We found that for high-angular momentum flow, the sonic point or sonic surface resides inside the ISCO radius. With the decrease in angular momentum, the sonic point or sonic surface can also exist outside the ISCO radius.
    
    \item Throughout our study, we did not find significant differences between magnetized and unmagnetized low angular momentum flows. All the timing and flow properties look quite similar in both cases. This is due to the fact that we restrict ourselves within the weak limits of a magnetic field. In our upcoming studies, we will explore strong magnetic field limits.
\end{itemize}

In light of the above findings, it is interesting to note that the intermediate angular momentum case is significantly different from other cases. This case has spiral filaments that are quite similar to those seen in MAD flows \citep[e.g,][]{Porth-etal2021,Begelman-etal2022} but they are seen slightly far from the black hole. Such spiral filaments are possible to connect the flaring activities in Sgr~A$^*$ \citep[e.g.,][]{Porth-etal2021,Ripperda-etal2022,Scepi-etal2022}. However, MAD flows are known to have very powerful jets, which is not clearly confirmed in the case for Sgr~A$^*$ \citep{Royster-etal2019,Yusef-Zadeh-etal2020}. Therefore, there are lots of ambiguities regarding the nature of the accretion flows around Sgr~A$^*$. Interestingly, in the intermediate angular momentum flows, we do not see any bipolar outflows or jets. At the same time, there is still uncertainty in explaining the observed light curve of Sgr~A$^*$ using the conventional SANE or MAD models \citep{Murchikova-Witzel2021,EHT2022,Wielgus-etal2022,Murchikova-etal2022}. Accordingly, we suggest that the intermediate angular momentum flow case could be a reasonable alternative to explaining the accretion flows around Sgr~A$^*$. We should mention that for such a study, we need to perform GRRT calculations, which we plan to do in our upcoming studies.

There have been lots of semi-analytical and numerical (axisymmetric, pseudo-Newtonian) studies that show low angular momentum flows can help explain the radiative (luminosity, spectra, etc.) and timing properties (quasi-periodic oscillations, flaring, etc.) in black hole X-ray binaries (BH-XRBs) as well as active galactic nuclei (AGNs), taking into account standing/oscillating shock solutions with multiple sonic points \citep[e.g.,][]{Chakrabarti-Titarchuk1995,Chakrabarti-etal2004,Das-etal2014,Aktar-etal2015,Chakrabarti2018,Dihingia-etal2018,Dihingia-etal2019a,Dihingia-etal2019b,Okuda-etal2019,Dihingia-etal2020,Okuda-etal2022}. However, with global 3D GRMHD simulations, it has never been tested. Due to magnetohydrodynamics, many rich phenomena may appear, such as reconnection-driven turbulence, MAD configurations, jets, magnetized winds, etc. \citep[e.g.,][]{Vourellis-etal2019,Nathanail-etal2019,Dihingia-etal2021,Ripperda-etal2022,Hong-Xuan-etal2023}. The semi-analytical expectation may or may not hold. To capture these phenomena, higher-resolution simulations than the current study are needed, which we plan to test in our future studies.

\begin{acknowledgments}
This research is supported by the National Natural Science Foundation of China (Grant No. 12273022), the Shanghai Municipality orientation program of Basic Research for International Scientists (grant no. 22JC1410600), and the National Key R\&D Program of China (No. 2023YFE0101200). The simulations were performed on the TDLI-Astro cluster in Tsung-Dao Lee Institute, Pi2.0, and Siyuan Mark-I clusters in the High-Performance Computing Center at Shanghai Jiao Tong University.
This work has made use of NASA's Astrophysics Data System (ADS). We appreciate the thoroughness and thoughtful comments provided by the anonymous reviewers that have improved the manuscript.
\end{acknowledgments}

\appendix
\section{Initial quantities}
\begin{figure*}
    \centering
    \includegraphics[scale=0.55]{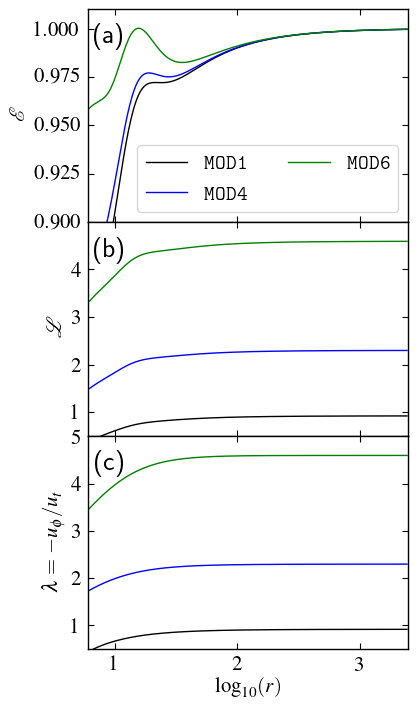}
    \includegraphics[scale=0.55]{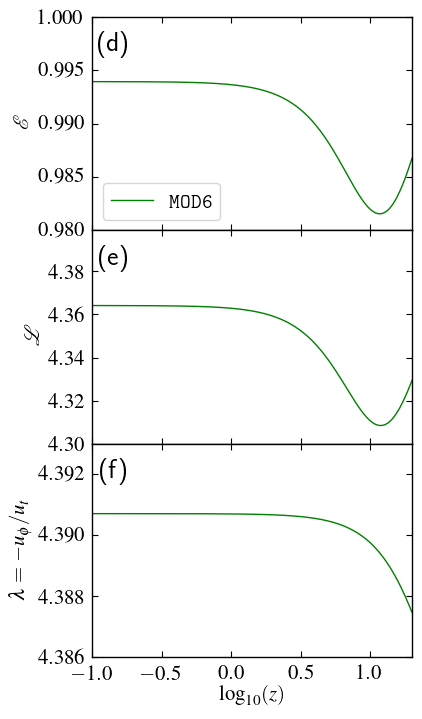}
    \caption{{\em Left}: Radial distribution ${\cal E}$, ${\cal L}$, and specific angular momentum $(\lambda)$ at the equatorial plan for different models before time evolution. {\em Right}: Same as {\em Left}, but vertical distribution at radius $r=20r_g$ only for \texttt{MOD6}.}
    \label{fig:a}
\end{figure*}

In this section, we show the radial and vertical distributions of some initial quantities, which are crucial to determining the evolution state of our simulation models. In Fig.~\ref{fig:a}, we show the radial distribution of ${\cal E}=-hu_t$, ${\cal L}=hu_\phi$, and specific angular momentum $(\lambda=-u_\phi/u_t)$ for different simulation models in panels (a), (b), and (c), respectively. In panels (d), (e), and (f), we show the same, but vertical distribution only for model \texttt{MOD6}. Since we follow the FM torus \cite{Fishbone-Moncrief1976} for the initial conditions, Accordingly, the profiles of ${\cal E}$, ${\cal L}$, and $\lambda$ are not constant throughout. The panels (a) and (d) suggest that flow is always less than unity $({\cal E}<1)$, that is, flow is gravitationally bound. The value of ${\cal L}$ scaled with the choice of ${\cal F}$. Finally, the specific angular momentum $(\lambda$) also has the same scaling as ${\cal L}$, and the maximum values of each case reach $\lambda_0$ as $r\gg1$.

In the {\em Right} panels of Fig.~\ref{fig:a}, we show a vertical distribution of the same quantities as the {\em Left} panels. For these panels, we only show a vertical distribution at radius $r=20r_g$ for model \texttt{MOD6}. We observe that the values of ${\cal E}$ and ${\cal L}$ decrease far from the equatorial plane initially but finally increase slightly. On the contrary, the value of specific angular momentum always decreases far from the equatorial plane (panel c). These features are similar at different radii and for other models. Accordingly, we do not show them here explicitly. 

\bibliography{references}{}

\begin{thebibliography}{}
\expandafter\ifx\csname natexlab\endcsname\relax\def\natexlab#1{#1}\fi
\providecommand{\url}[1]{\href{#1}{#1}}
\providecommand{\dodoi}[1]{doi:~\href{http://doi.org/#1}{\nolinkurl{#1}}}
\providecommand{\doeprint}[1]{\href{http://ascl.net/#1}{\nolinkurl{http://ascl.net/#1}}}
\providecommand{\doarXiv}[1]{\href{https://arxiv.org/abs/#1}{\nolinkurl{https://arxiv.org/abs/#1}}}

\bibitem[{{Abramowicz} {et~al.}(1983){Abramowicz}, {Calvani}, \& {Nobili}}]{Abramowicz-etal1983}
{Abramowicz}, M.~A., {Calvani}, M., \& {Nobili}, L. 1983, \nat, 302, 597, \dodoi{10.1038/302597a0}

\bibitem[{{Abramowicz} {et~al.}(1988){Abramowicz}, {Czerny}, {Lasota}, \& {Szuszkiewicz}}]{Abramowicz-etal1988}
{Abramowicz}, M.~A., {Czerny}, B., {Lasota}, J.~P., \& {Szuszkiewicz}, E. 1988, \apj, 332, 646, \dodoi{10.1086/166683}

\bibitem[{{Aktar} {et~al.}(2015){Aktar}, {Das}, \& {Nandi}}]{Aktar-etal2015}
{Aktar}, R., {Das}, S., \& {Nandi}, A. 2015, \mnras, 453, 3414, \dodoi{10.1093/mnras/stv1874}

\bibitem[{{Balbus} \& {Hawley}(1991)}]{Balbus-Hawley1991}
{Balbus}, S.~A., \& {Hawley}, J.~F. 1991, \apj, 376, 214, \dodoi{10.1086/170270}

\bibitem[{{Balbus} \& {Hawley}(1998)}]{Balbus-Hawley1998}
---. 1998, Reviews of Modern Physics, 70, 1, \dodoi{10.1103/RevModPhys.70.1}

\bibitem[{{Begelman} {et~al.}(2022){Begelman}, {Scepi}, \& {Dexter}}]{Begelman-etal2022}
{Begelman}, M.~C., {Scepi}, N., \& {Dexter}, J. 2022, \mnras, 511, 2040, \dodoi{10.1093/mnras/stab3790}

\bibitem[{{Blandford} \& {Begelman}(2004)}]{Blandford-Begelman2004}
{Blandford}, R.~D., \& {Begelman}, M.~C. 2004, \mnras, 349, 68, \dodoi{10.1111/j.1365-2966.2004.07425.x}

\bibitem[{{Bondi}(1952)}]{Bondi1952}
{Bondi}, H. 1952, \mnras, 112, 195, \dodoi{10.1093/mnras/112.2.195}

\bibitem[{{Chakrabarti} \& {Titarchuk}(1995)}]{Chakrabarti-Titarchuk1995}
{Chakrabarti}, S., \& {Titarchuk}, L.~G. 1995, \apj, 455, 623, \dodoi{10.1086/176610}

\bibitem[{{Chakrabarti}(1989)}]{Chakrabarti1989}
{Chakrabarti}, S.~K. 1989, \apj, 347, 365, \dodoi{10.1086/168125}

\bibitem[{{Chakrabarti}(1996)}]{Chakrabarti1996}
---. 1996, \mnras, 283, 325, \dodoi{10.1093/mnras/283.1.325}

\bibitem[{{Chakrabarti}(2015)}]{Chakrabarti-etal2015}
{Chakrabarti}, S.~K. 2015, in Astronomical Society of India Conference Series, Vol.~12, Astronomical Society of India Conference Series, 9--16.
\newblock \doarXiv{1509.00565}

\bibitem[{{Chakrabarti}(2018)}]{Chakrabarti2018}
{Chakrabarti}, S.~K. 2018, in Fourteenth Marcel Grossmann Meeting - MG14, ed. M.~{Bianchi}, R.~T. {Jansen}, \& R.~{Ruffini}, 369--384, \dodoi{10.1142/9789813226609_0020}

\bibitem[{{Chakrabarti} {et~al.}(2004){Chakrabarti}, {Acharyya}, \& {Molteni}}]{Chakrabarti-etal2004}
{Chakrabarti}, S.~K., {Acharyya}, K., \& {Molteni}, D. 2004, \aap, 421, 1, \dodoi{10.1051/0004-6361:20034523}

\bibitem[{{Cuadra} {et~al.}(2008){Cuadra}, {Nayakshin}, \& {Martins}}]{Cuadra-etal2008}
{Cuadra}, J., {Nayakshin}, S., \& {Martins}, F. 2008, \mnras, 383, 458, \dodoi{10.1111/j.1365-2966.2007.12573.x}

\bibitem[{{Das}(2007)}]{Das2007}
{Das}, S. 2007, \mnras, 376, 1659, \dodoi{10.1111/j.1365-2966.2007.11501.x}

\bibitem[{{Das} {et~al.}(2014){Das}, {Chattopadhyay}, {Nandi}, \& {Molteni}}]{Das-etal2014}
{Das}, S., {Chattopadhyay}, I., {Nandi}, A., \& {Molteni}, D. 2014, \mnras, 442, 251, \dodoi{10.1093/mnras/stu864}

\bibitem[{{Davis} \& {Tchekhovskoy}(2020)}]{Davis-Tchekhovskoy2020}
{Davis}, S.~W., \& {Tchekhovskoy}, A. 2020, \araa, 58, annurev, \dodoi{10.1146/annurev-astro-081817-051905}

\bibitem[{{Dexter} {et~al.}(2020){Dexter}, {Tchekhovskoy}, {Jim{\'e}nez-Rosales}, {Ressler}, {Baub{\"o}ck}, {Dallilar}, {de Zeeuw}, {Eisenhauer}, {von Fellenberg}, {Gao}, {Genzel}, {Gillessen}, {Habibi}, {Ott}, {Stadler}, {Straub}, \& {Widmann}}]{Dexter-etal2020}
{Dexter}, J., {Tchekhovskoy}, A., {Jim{\'e}nez-Rosales}, A., {et~al.} 2020, \mnras, 497, 4999, \dodoi{10.1093/mnras/staa2288}

\bibitem[{{Dihingia} {et~al.}(2018{\natexlab{a}}){Dihingia}, {Das}, {Maity}, \& {Chakrabarti}}]{Dihingia-etal2018a}
{Dihingia}, I.~K., {Das}, S., {Maity}, D., \& {Chakrabarti}, S. 2018{\natexlab{a}}, \prd, 98, 083004, \dodoi{10.1103/PhysRevD.98.083004}

\bibitem[{{Dihingia} {et~al.}(2019{\natexlab{a}}){Dihingia}, {Das}, {Maity}, \& {Nandi}}]{Dihingia-etal2019b}
{Dihingia}, I.~K., {Das}, S., {Maity}, D., \& {Nandi}, A. 2019{\natexlab{a}}, Mon. Not. Roy. Astron. Soc., 1852, \dodoi{10.1093/mnras/stz1933}

\bibitem[{{Dihingia} {et~al.}(2018{\natexlab{b}}){Dihingia}, {Das}, \& {Mandal}}]{Dihingia-etal2018}
{Dihingia}, I.~K., {Das}, S., \& {Mandal}, S. 2018{\natexlab{b}}, \mnras, 475, 2164, \dodoi{10.1093/mnras/stx3269}

\bibitem[{{Dihingia} {et~al.}(2019{\natexlab{b}}){Dihingia}, {Das}, \& {Nandi}}]{Dihingia-etal2019a}
{Dihingia}, I.~K., {Das}, S., \& {Nandi}, A. 2019{\natexlab{b}}, Mon. Not. Roy. Astron. Soc., 484, 3209, \dodoi{10.1093/mnras/stz168}

\bibitem[{{Dihingia} {et~al.}(2020){Dihingia}, {Das}, {Prabhakar}, \& {Mand al}}]{Dihingia-etal2020}
{Dihingia}, I.~K., {Das}, S., {Prabhakar}, G., \& {Mand al}, S. 2020, \mnras, 496, 3043, \dodoi{10.1093/mnras/staa1687}

\bibitem[{{Dihingia} {et~al.}(2021){Dihingia}, {Vaidya}, \& {Fendt}}]{Dihingia-etal2021}
{Dihingia}, I.~K., {Vaidya}, B., \& {Fendt}, C. 2021, \mnras, 505, 3596, \dodoi{10.1093/mnras/stab1512}

\bibitem[{{Dihingia} {et~al.}(2022){Dihingia}, {Vaidya}, \& {Fendt}}]{Dihingia-etal2022}
---. 2022, \mnras, 517, 5032, \dodoi{10.1093/mnras/stac3021}

\bibitem[{{Event Horizon Telescope Collaboration} {et~al.}(2019){Event Horizon Telescope Collaboration}, {Akiyama}, {Alberdi}, {Alef}, {Asada}, {Azulay}, {Baczko}, {Ball}, {Balokovi{\'c}}, {Barrett}, {Bintley}, {Blackburn}, {Boland}, {Bouman}, {Bower}, {Bremer}, {Brinkerink}, {Brissenden}, {Britzen}, {Broderick}, {Broguiere}, {Bronzwaer}, {Byun}, {Carlstrom}, {Chael}, {Chan}, {Chatterjee}, {Chatterjee}, {Chen}, {Chen}, {Cho}, {Christian}, {Conway}, {Cordes}, {Crew}, {Cui}, {Davelaar}, {De Laurentis}, {Deane}, {Dempsey}, {Desvignes}, {Dexter}, {Doeleman}, {Eatough}, {Falcke}, {Fish}, {Fomalont}, {Fraga-Encinas}, {Freeman}, {Friberg}, {Fromm}, {G{\'o}mez}, {Galison}, {Gammie}, {Garc{\'\i}a}, {Gentaz}, {Georgiev}, {Goddi}, {Gold}, {Gu}, {Gurwell}, {Hada}, {Hecht}, {Hesper}, {Ho}, {Ho}, {Honma}, {Huang}, {Huang}, {Hughes}, {Ikeda}, {Inoue}, {Issaoun}, {James}, {Jannuzi}, {Janssen}, {Jeter}, {Jiang}, {Johnson}, {Jorstad}, {Jung}, {Karami}, {Karuppusamy}, {Kawashima}, {Keating}, {Kettenis}, {Kim}, {Kim}, {Kim},
  {Kino}, {Koay}, {Koch}, {Koyama}, {Kramer}, {Kramer}, {Krichbaum}, {Kuo}, {Lauer}, {Lee}, {Li}, {Li}, {Lindqvist}, {Liu}, {Liuzzo}, {Lo}, {Lobanov}, {Loinard}, {Lonsdale}, {Lu}, {MacDonald}, {Mao}, {Markoff}, {Marrone}, {Marscher}, {Mart{\'\i}-Vidal}, {Matsushita}, {Matthews}, {Medeiros}, {Menten}, {Mizuno}, {Mizuno}, {Moran}, {Moriyama}, {Moscibrodzka}, {M{\"u}ller}, {Nagai}, {Nagar}, {Nakamura}, {Narayan}, {Narayanan}, {Natarajan}, {Neri}, {Ni}, {Noutsos}, {Okino}, {Olivares}, {Ortiz-Le{\'o}n}, {Oyama}, {{\"O}zel}, {Palumbo}, {Patel}, {Pen}, {Pesce}, {Pi{\'e}tu}, {Plambeck}, {PopStefanija}, {Porth}, {Prather}, {Preciado-L{\'o}pez}, {Psaltis}, {Pu}, {Ramakrishnan}, {Rao}, {Rawlings}, {Raymond}, {Rezzolla}, {Ripperda}, {Roelofs}, {Rogers}, {Ros}, {Rose}, {Roshanineshat}, {Rottmann}, {Roy}, {Ruszczyk}, {Ryan}, {Rygl}, {S{\'a}nchez}, {S{\'a}nchez-Arguelles}, {Sasada}, {Savolainen}, {Schloerb}, {Schuster}, {Shao}, {Shen}, {Small}, {Sohn}, {SooHoo}, {Tazaki}, {Tiede}, {Tilanus}, {Titus}, {Toma}, {Torne},
  {Trent}, {Trippe}, {Tsuda}, {van Bemmel}, {van Langevelde}, {van Rossum}, {Wagner}, {Wardle}, {Weintroub}, {Wex}, {Wharton}, {Wielgus}, {Wong}, {Wu}, {Young}, {Young}, {Younsi}, {Yuan}, {Yuan}, {Zensus}, {Zhao}, {Zhao}, {Zhu}, {Algaba}, {Allardi}, {Amestica}, {Anczarski}, {Bach}, {Baganoff}, {Beaudoin}, {Benson}, {Berthold}, {Blanchard}, {Blundell}, {Bustamente}, {Cappallo}, {Castillo-Dom{\'\i}nguez}, {Chang}, {Chang}, {Chang}, {Chen}, {Chilson}, {Chuter}, {C{\'o}rdova Rosado}, {Coulson}, {Crawford}, {Crowley}, {David}, {Derome}, {Dexter}, {Dornbusch}, {Dudevoir}, {Dzib}, {Eckart}, {Eckert}, {Erickson}, {Everett}, {Faber}, {Farah}, {Fath}, {Folkers}, {Forbes}, {Freund}, {G{\'o}mez-Ruiz}, {Gale}, {Gao}, {Geertsema}, {Graham}, {Greer}, {Grosslein}, {Gueth}, {Haggard}, {Halverson}, {Han}, {Han}, {Hao}, {Hasegawa}, {Henning}, {Hern{\'a}ndez-G{\'o}mez}, {Herrero-Illana}, {Heyminck}, {Hirota}, {Hoge}, {Huang}, {Impellizzeri}, {Jiang}, {Kamble}, {Keisler}, {Kimura}, {Kono}, {Kubo}, {Kuroda}, {Lacasse}, {Laing},
  {Leitch}, {Li}, {Lin}, {Liu}, {Liu}, {Lu}, {Marson}, {Martin-Cocher}, {Massingill}, {Matulonis}, {McColl}, {McWhirter}, {Messias}, {Meyer-Zhao}, {Michalik}, {Monta{\~n}a}, {Montgomerie}, {Mora-Klein}, {Muders}, {Nadolski}, {Navarro}, {Neilsen}, {Nguyen}, {Nishioka}, {Norton}, {Nowak}, {Nystrom}, {Ogawa}, {Oshiro}, {Oyama}, {Parsons}, {Paine}, {Pe{\~n}alver}, {Phillips}, {Poirier}, {Pradel}, {Primiani}, {Raffin}, {Rahlin}, {Reiland}, {Risacher}, {Ruiz}, {S{\'a}ez-Mada{\'\i}n}, {Sassella}, {Schellart}, {Shaw}, {Silva}, {Shiokawa}, {Smith}, {Snow}, {Souccar}, {Sousa}, {Sridharan}, {Srinivasan}, {Stahm}, {Stark}, {Story}, {Timmer}, {Vertatschitsch}, {Walther}, {Wei}, {Whitehorn}, {Whitney}, {Woody}, {Wouterloot}, {Wright}, {Yamaguchi}, {Yu}, {Zeballos}, {Zhang}, \& {Ziurys}}]{EHTI-2019}
{Event Horizon Telescope Collaboration}, {Akiyama}, K., {Alberdi}, A., {et~al.} 2019, \apjl, 875, L1, \dodoi{10.3847/2041-8213/ab0ec7}

\bibitem[{{Event Horizon Telescope Collaboration} {et~al.}(2021){Event Horizon Telescope Collaboration}, {Akiyama}, {Algaba}, {Alberdi}, {Alef}, {Anantua}, {Asada}, {Azulay}, {Baczko}, {Ball}, {Balokovi{\'c}}, {Barrett}, {Benson}, {Bintley}, {Blackburn}, {Blundell}, {Boland}, {Bouman}, {Bower}, {Boyce}, {Bremer}, {Brinkerink}, {Brissenden}, {Britzen}, {Broderick}, {Broguiere}, {Bronzwaer}, {Byun}, {Carlstrom}, {Chael}, {Chan}, {Chatterjee}, {Chatterjee}, {Chen}, {Chen}, {Chesler}, {Cho}, {Christian}, {Conway}, {Cordes}, {Crawford}, {Crew}, {Cruz-Osorio}, {Cui}, {Davelaar}, {De Laurentis}, {Deane}, {Dempsey}, {Desvignes}, {Dexter}, {Doeleman}, {Eatough}, {Falcke}, {Farah}, {Fish}, {Fomalont}, {Ford}, {Fraga-Encinas}, {Freeman}, {Friberg}, {Fromm}, {Fuentes}, {Galison}, {Gammie}, {Garc{\'\i}a}, {Gentaz}, {Georgiev}, {Goddi}, {Gold}, {G{\'o}mez}, {G{\'o}mez-Ruiz}, {Gu}, {Gurwell}, {Hada}, {Haggard}, {Hecht}, {Hesper}, {Ho}, {Ho}, {Honma}, {Huang}, {Huang}, {Hughes}, {Ikeda}, {Inoue}, {Issaoun}, {James},
  {Jannuzi}, {Janssen}, {Jeter}, {Jiang}, {Jimenez-Rosales}, {Johnson}, {Jorstad}, {Jung}, {Karami}, {Karuppusamy}, {Kawashima}, {Keating}, {Kettenis}, {Kim}, {Kim}, {Kim}, {Kim}, {Kino}, {Koay}, {Kofuji}, {Koch}, {Koyama}, {Kramer}, {Kramer}, {Krichbaum}, {Kuo}, {Lauer}, {Lee}, {Levis}, {Li}, {Li}, {Lindqvist}, {Lico}, {Lindahl}, {Liu}, {Liu}, {Liuzzo}, {Lo}, {Lobanov}, {Loinard}, {Lonsdale}, {Lu}, {MacDonald}, {Mao}, {Marchili}, {Markoff}, {Marrone}, {Marscher}, {Mart{\'\i}-Vidal}, {Matsushita}, {Matthews}, {Medeiros}, {Menten}, {Mizuno}, {Mizuno}, {Moran}, {Moriyama}, {Moscibrodzka}, {M{\"u}ller}, {Musoke}, {Mej{\'\i}as}, {Michalik}, {Nadolski}, {Nagai}, {Nagar}, {Nakamura}, {Narayan}, {Narayanan}, {Natarajan}, {Nathanail}, {Neilsen}, {Neri}, {Ni}, {Noutsos}, {Nowak}, {Okino}, {Olivares}, {Ortiz-Le{\'o}n}, {Oyama}, {{\"O}zel}, {Palumbo}, {Park}, {Patel}, {Pen}, {Pesce}, {Pi{\'e}tu}, {Plambeck}, {PopStefanija}, {Porth}, {P{\"o}tzl}, {Prather}, {Preciado-L{\'o}pez}, {Psaltis}, {Pu}, {Ramakrishnan}, {Rao},
  {Rawlings}, {Raymond}, {Rezzolla}, {Ricarte}, {Ripperda}, {Roelofs}, {Rogers}, {Ros}, {Rose}, {Roshanineshat}, {Rottmann}, {Roy}, {Ruszczyk}, {Rygl}, {S{\'a}nchez}, {S{\'a}nchez-Arguelles}, {Sasada}, {Savolainen}, {Schloerb}, {Schuster}, {Shao}, {Shen}, {Small}, {Sohn}, {SooHoo}, {Sun}, {Tazaki}, {Tetarenko}, {Tiede}, {Tilanus}, {Titus}, {Toma}, {Torne}, {Trent}, {Traianou}, {Trippe}, {van Bemmel}, {van Langevelde}, {van Rossum}, {Wagner}, {Ward-Thompson}, {Wardle}, {Weintroub}, {Wex}, {Wharton}, {Wielgus}, {Wong}, {Wu}, {Yoon}, {Young}, {Young}, {Younsi}, {Yuan}, {Yuan}, {Zensus}, {Zhao}, \& {Zhao}}]{EHTVII-etal2021}
{Event Horizon Telescope Collaboration}, {Akiyama}, K., {Algaba}, J.~C., {et~al.} 2021, \apjl, 910, L12, \dodoi{10.3847/2041-8213/abe71d}

\bibitem[{{Event Horizon Telescope Collaboration} {et~al.}(2022){Event Horizon Telescope Collaboration}, {Akiyama}, {Alberdi}, {Alef}, {Algaba}, {Anantua}, {Asada}, {Azulay}, {Bach}, {Baczko}, {Ball}, {Balokovi{\'c}}, {Barrett}, {Baub{\"o}ck}, {Benson}, {Bintley}, {Blackburn}, {Blundell}, {Bouman}, {Bower}, {Boyce}, {Bremer}, {Brinkerink}, {Brissenden}, {Britzen}, {Broderick}, {Broguiere}, {Bronzwaer}, {Bustamante}, {Byun}, {Carlstrom}, {Ceccobello}, {Chael}, {Chan}, {Chatterjee}, {Chatterjee}, {Chen}, {Chen}, {Cheng}, {Cho}, {Christian}, {Conroy}, {Conway}, {Cordes}, {Crawford}, {Crew}, {Cruz-Osorio}, {Cui}, {Davelaar}, {De Laurentis}, {Deane}, {Dempsey}, {Desvignes}, {Dexter}, {Dhruv}, {Doeleman}, {Dougal}, {Dzib}, {Eatough}, {Emami}, {Falcke}, {Farah}, {Fish}, {Fomalont}, {Ford}, {Fraga-Encinas}, {Freeman}, {Friberg}, {Fromm}, {Fuentes}, {Galison}, {Gammie}, {Garc{\'\i}a}, {Gentaz}, {Georgiev}, {Goddi}, {Gold}, {G{\'o}mez-Ruiz}, {G{\'o}mez}, {Gu}, {Gurwell}, {Hada}, {Haggard}, {Haworth}, {Hecht}, {Hesper},
  {Heumann}, {Ho}, {Ho}, {Honma}, {Huang}, {Huang}, {Hughes}, {Ikeda}, {Violette Impellizzeri}, {Inoue}, {Issaoun}, {James}, {Jannuzi}, {Janssen}, {Jeter}, {Jiang}, {Jim{\'e}nez-Rosales}, {Johnson}, {Jorstad}, {Joshi}, {Jung}, {Karami}, {Karuppusamy}, {Kawashima}, {Keating}, {Kettenis}, {Kim}, {Kim}, {Kim}, {Kim}, {Kino}, {Koay}, {Kocherlakota}, {Kofuji}, {Koch}, {Koyama}, {Kramer}, {Kramer}, {Krichbaum}, {Kuo}, {La Bella}, {Lauer}, {Lee}, {Lee}, {Leung}, {Levis}, {Li}, {Lico}, {Lindahl}, {Lindqvist}, {Lisakov}, {Liu}, {Liu}, {Liuzzo}, {Lo}, {Lobanov}, {Loinard}, {Lonsdale}, {Lu}, {Mao}, {Marchili}, {Markoff}, {Marrone}, {Marscher}, {Mart{\'\i}-Vidal}, {Matsushita}, {Matthews}, {Medeiros}, {Menten}, {Michalik}, {Mizuno}, {Mizuno}, {Moran}, {Moriyama}, {Moscibrodzka}, {M{\"u}ller}, {Mus}, {Musoke}, {Myserlis}, {Nadolski}, {Nagai}, {Nagar}, {Nakamura}, {Narayan}, {Narayanan}, {Natarajan}, {Nathanail}, {Navarro Fuentes}, {Neilsen}, {Neri}, {Ni}, {Noutsos}, {Nowak}, {Oh}, {Okino}, {Olivares}, {Ortiz-Le{\'o}n},
  {Oyama}, {{\"O}zel}, {Palumbo}, {Filippos Paraschos}, {Park}, {Parsons}, {Patel}, {Pen}, {Pesce}, {Pi{\'e}tu}, {Plambeck}, {PopStefanija}, {Porth}, {P{\"o}tzl}, {Prather}, {Preciado-L{\'o}pez}, {Psaltis}, {Pu}, {Ramakrishnan}, {Rao}, {Rawlings}, {Raymond}, {Rezzolla}, {Ricarte}, {Ripperda}, {Roelofs}, {Rogers}, {Ros}, {Romero-Ca{\~n}izales}, {Roshanineshat}, {Rottmann}, {Roy}, {Ruiz}, {Ruszczyk}, {Rygl}, {S{\'a}nchez}, {S{\'a}nchez-Arg{\"u}elles}, {S{\'a}nchez-Portal}, {Sasada}, {Satapathy}, {Savolainen}, {Schloerb}, {Schonfeld}, {Schuster}, {Shao}, {Shen}, {Small}, {Sohn}, {SooHoo}, {Souccar}, {Sun}, {Tazaki}, {Tetarenko}, {Tiede}, {Tilanus}, {Titus}, {Torne}, {Traianou}, {Trent}, {Trippe}, {Turk}, {van Bemmel}, {van Langevelde}, {van Rossum}, {Vos}, {Wagner}, {Ward-Thompson}, {Wardle}, {Weintroub}, {Wex}, {Wharton}, {Wielgus}, {Wiik}, {Witzel}, {Wondrak}, {Wong}, {Wu}, {Yamaguchi}, {Yoon}, {Young}, {Young}, {Younsi}, {Yuan}, {Yuan}, {Zensus}, {Zhang}, {Zhao}, {Zhao}, {Chan}, {Qiu}, {Ressler}, \&
  {White}}]{EHT2022}
{Event Horizon Telescope Collaboration}, {Akiyama}, K., {Alberdi}, A., {et~al.} 2022, \apjl, 930, L16, \dodoi{10.3847/2041-8213/ac6672}

\bibitem[{{Event Horizon Telescope Collaboration} {et~al.}(2024){Event Horizon Telescope Collaboration}, {Akiyama}, {Alberdi}, {Alef}, {Algaba}, {Anantua}, {Asada}, {Azulay}, {Bach}, {Baczko}, {Ball}, {Balokovi{\'c}}, {Bandyopadhyay}, {Barrett}, {Baub{\"o}ck}, {Benson}, {Bintley}, {Blackburn}, {Blundell}, {Bouman}, {Bower}, {Boyce}, {Bremer}, {Brissenden}, {Britzen}, {Broderick}, {Broguiere}, {Bronzwaer}, {Bustamante}, {Carlstrom}, {Chael}, {Chan}, {Chang}, {Chatterjee}, {Chatterjee}, {Chen}, {Chen}, {Cheng}, {Cho}, {Christian}, {Conroy}, {Conway}, {Crawford}, {Crew}, {Cruz-Osorio}, {Cui}, {Dahale}, {Davelaar}, {De Laurentis}, {Deane}, {Dempsey}, {Desvignes}, {Dexter}, {Dhruv}, {Dihingia}, {Doeleman}, {Dzib}, {Eatough}, {Emami}, {Falcke}, {Farah}, {Fish}, {Fomalont}, {Ford}, {Foschi}, {Fraga-Encinas}, {Freeman}, {Friberg}, {Fromm}, {Fuentes}, {Galison}, {Gammie}, {Garc{\'\i}a}, {Gentaz}, {Georgiev}, {Goddi}, {Gold}, {G{\'o}mez-Ruiz}, {G{\'o}mez}, {Gu}, {Gurwell}, {Hada}, {Haggard}, {Hesper}, {Heumann}, {Ho},
  {Ho}, {Honma}, {Huang}, {Huang}, {Hughes}, {Ikeda}, {Violette Impellizzeri}, {Inoue}, {Issaoun}, {James}, {Jannuzi}, {Janssen}, {Jeter}, {Jiang}, {Jim{\'e}nez-Rosales}, {Johnson}, {Jorstad}, {Jones}, {Joshi}, {Jung}, {Karuppusamy}, {Kawashima}, {Keating}, {Kettenis}, {Kim}, {Kim}, {Kim}, {Kim}, {Kino}, {Koay}, {Kocherlakota}, {Kofuji}, {Koch}, {Koyama}, {Kramer}, {Kramer}, {Kramer}, {Krichbaum}, {Kuo}, {La Bella}, {Lee}, {Levis}, {Li}, {Lico}, {Lindahl}, {Lindqvist}, {Lisakov}, {Liu}, {Liu}, {Liuzzo}, {Lo}, {Lobanov}, {Loinard}, {Lonsdale}, {Lowitz}, {Lu}, {MacDonald}, {Mao}, {Marchili}, {Markoff}, {Marrone}, {Marscher}, {Mart{\'\i}-Vidal}, {Matsushita}, {Matthews}, {Medeiros}, {Menten}, {Mizuno}, {Mizuno}, {Montgomery}, {Moran}, {Moriyama}, {Moscibrodzka}, {Mulaudzi}, {M{\"u}ller}, {M{\"u}ller}, {Mus}, {Musoke}, {Myserlis}, {Nagai}, {Nagar}, {Nakamura}, {Narayanan}, {Natarajan}, {Nathanail}, {Fuentes}, {Neilsen}, {Ni}, {Nowak}, {Oh}, {Okino}, {Olivares}, {Oyama}, {{\"O}zel}, {Palumbo}, {Paraschos}, {Park},
  {Parsons}, {Patel}, {Pen}, {Pesce}, {Pi{\'e}tu}, {PopStefanija}, {Porth}, {Prather}, {Psaltis}, {Pu}, {Ramakrishnan}, {Rao}, {Rawlings}, {Raymond}, {Rezzolla}, {Ricarte}, {Ripperda}, {Roelofs}, {Romero-Ca{\~n}izales}, {Ros}, {Roshanineshat}, {Rottmann}, {Roy}, {Ruiz}, {Ruszczyk}, {Rygl}, {S{\'a}nchez}, {S{\'a}nchez-Arg{\"u}elles}, {S{\'a}nchez-Portal}, {Sasada}, {Satapathy}, {Savolainen}, {Schloerb}, {Schonfeld}, {Schuster}, {Shao}, {Shen}, {Small}, {Sohn}, {SooHoo}, {Salas}, {Souccar}, {Stanway}, {Sun}, {Tazaki}, {Tetarenko}, {Tiede}, {Tilanus}, {Titus}, {Toma}, {Torne}, {Toscano}, {Traianou}, {Trent}, {Trippe}, {Turk}, {van Bemmel}, {van Langevelde}, {van Rossum}, {Vos}, {Wagner}, {Ward-Thompson}, {Wardle}, {Washington}, {Weintroub}, {Wharton}, {Wielgus}, {Wiik}, {Witzel}, {Wondrak}, {Wong}, {Wu}, {Yadlapalli}, {Yamaguchi}, {Yfantis}, {Yoon}, {Young}, {Younsi}, {Yu}, {Yuan}, {Yuan}, {Anton Zensus}, {Zhang}, {Zhao}, {Zhao}, {Allardi}, {Chang}, {Chang}, {Chang}, {Chen}, {Chilson}, {Faber}, {Gale}, {Han},
  {Han}, {Hasegawa}, {Hern{\'a}ndez-Rebollar}, {Huang}, {Jiang}, {Jinchi}, {Kimura}, {Kubo}, {Li}, {Lin}, {Liu}, {Liu}, {Lu}, {Martin-Cocher}, {Meyer-Zhao}, {Monta{\~n}a}, {Moraghan}, {Moreno-Nolasco}, {Nishioka}, {Norton}, {Nystrom}, {Ogawa}, {Oshiro}, {Pradel}, {Principe}, {Raffin}, {Rodr{\'\i}guez-Montoya}, {Shaw}, {Snow}, {Sridharan}, {Srinivasan}, {Wei}, \& {Yu}}]{EHT2024}
---. 2024, \aap, 681, A79, \dodoi{10.1051/0004-6361/202347932}

\bibitem[{{Fishbone} \& {Moncrief}(1976)}]{Fishbone-Moncrief1976}
{Fishbone}, L.~G., \& {Moncrief}, V. 1976, \apj, 207, 962, \dodoi{10.1086/154565}

\bibitem[{{Frank} {et~al.}(2002){Frank}, {King}, \& {Raine}}]{Frank-etal2002}
{Frank}, J., {King}, A., \& {Raine}, D.~J. 2002, {Accretion Power in Astrophysics: Third Edition} (Cambridge, UK: Cambridge University Press, February 2002)

\bibitem[{{Fukue}(1987)}]{Fukue1987}
{Fukue}, J. 1987, \pasj, 39, 309

\bibitem[{{Giri} {et~al.}(2010){Giri}, {Chakrabarti}, {Samanta}, \& {Ryu}}]{Giri-etal2010}
{Giri}, K., {Chakrabarti}, S.~K., {Samanta}, M.~M., \& {Ryu}, D. 2010, \mnras, 403, 516, \dodoi{10.1111/j.1365-2966.2009.16147.x}

\bibitem[{{Hawley} {et~al.}(1984{\natexlab{a}}){Hawley}, {Smarr}, \& {Wilson}}]{Hawley-etal1984a}
{Hawley}, J.~F., {Smarr}, L.~L., \& {Wilson}, J.~R. 1984{\natexlab{a}}, \apjs, 55, 211, \dodoi{10.1086/190953}

\bibitem[{{Hawley} {et~al.}(1984{\natexlab{b}}){Hawley}, {Smarr}, \& {Wilson}}]{Hawley-etal1984b}
---. 1984{\natexlab{b}}, \apj, 277, 296, \dodoi{10.1086/161696}

\bibitem[{{Jiang} {et~al.}(2023){Jiang}, {Mizuno}, {Fromm}, \& {Nathanail}}]{Hong-Xuan-etal2023}
{Jiang}, H.-X., {Mizuno}, Y., {Fromm}, C.~M., \& {Nathanail}, A. 2023, \mnras, 522, 2307, \dodoi{10.1093/mnras/stad1106}

\bibitem[{{Kim} {et~al.}(2017){Kim}, {Garain}, {Balsara}, \& {Chakrabarti}}]{Kim-etal2017}
{Kim}, J., {Garain}, S.~K., {Balsara}, D.~S., \& {Chakrabarti}, S.~K. 2017, \mnras, 472, 542, \dodoi{10.1093/mnras/stx1986}

\bibitem[{{Kim} {et~al.}(2019){Kim}, {Garain}, {Chakrabarti}, \& {Balsara}}]{Kim-etal2019}
{Kim}, J., {Garain}, S.~K., {Chakrabarti}, S.~K., \& {Balsara}, D.~S. 2019, \mnras, 482, 3636, \dodoi{10.1093/mnras/sty2953}

\bibitem[{{Korobkin} {et~al.}(2013){Korobkin}, {Abdikamalov}, {Stergioulas}, {Schnetter}, {Zink}, {Rosswog}, \& {Ott}}]{Korobkin-etal2013}
{Korobkin}, O., {Abdikamalov}, E., {Stergioulas}, N., {et~al.} 2013, \mnras, 431, 349, \dodoi{10.1093/mnras/stt166}

\bibitem[{{Kumar} \& {Chattopadhyay}(2017)}]{Kumar-Chattopadhyay2017}
{Kumar}, R., \& {Chattopadhyay}, I. 2017, \mnras, 469, 4221, \dodoi{10.1093/mnras/stx1091}

\bibitem[{{Lanzafame} {et~al.}(1998){Lanzafame}, {Molteni}, \& {Chakrabarti}}]{Lanzafame-etal1998}
{Lanzafame}, G., {Molteni}, D., \& {Chakrabarti}, S.~K. 1998, \mnras, 299, 799, \dodoi{10.1046/j.1365-8711.1998.01816.x}

\bibitem[{{Masuda} \& {Eriguchi}(1997)}]{Masuda-Eriguchi1997}
{Masuda}, N., \& {Eriguchi}, Y. 1997, \apj, 489, 804, \dodoi{10.1086/304818}

\bibitem[{{Mitra} {et~al.}(2022){Mitra}, {Maity}, {Dihingia}, \& {Das}}]{Mitra-etal2022}
{Mitra}, S., {Maity}, D., {Dihingia}, I.~K., \& {Das}, S. 2022, \mnras, 516, 5092, \dodoi{10.1093/mnras/stac2431}

\bibitem[{{Mizuno}(2022)}]{Mizuno2022}
{Mizuno}, Y. 2022, Universe, 8, 85, \dodoi{10.3390/universe8020085}

\bibitem[{{Mizuno} {et~al.}(2021){Mizuno}, {Fromm}, {Younsi}, {Porth}, {Olivares}, \& {Rezzolla}}]{Mizuno-etal2021}
{Mizuno}, Y., {Fromm}, C.~M., {Younsi}, Z., {et~al.} 2021, \mnras, 506, 741, \dodoi{10.1093/mnras/stab1753}

\bibitem[{{Molteni} {et~al.}(1994){Molteni}, {Lanzafame}, \& {Chakrabarti}}]{Molteni-etal1994}
{Molteni}, D., {Lanzafame}, G., \& {Chakrabarti}, S.~K. 1994, \apj, 425, 161, \dodoi{10.1086/173972}

\bibitem[{{Molteni} {et~al.}(1996{\natexlab{a}}){Molteni}, {Ryu}, \& {Chakrabarti}}]{Molteni-etal1996a}
{Molteni}, D., {Ryu}, D., \& {Chakrabarti}, S.~K. 1996{\natexlab{a}}, \apj, 470, 460, \dodoi{10.1086/177877}

\bibitem[{{Molteni} {et~al.}(1996{\natexlab{b}}){Molteni}, {Sponholz}, \& {Chakrabarti}}]{Molteni-etal1996b}
{Molteni}, D., {Sponholz}, H., \& {Chakrabarti}, S.~K. 1996{\natexlab{b}}, \apj, 457, 805, \dodoi{10.1086/176775}

\bibitem[{{Murchikova} {et~al.}(2022){Murchikova}, {White}, \& {Ressler}}]{Murchikova-etal2022}
{Murchikova}, L., {White}, C.~J., \& {Ressler}, S.~M. 2022, \apjl, 932, L21, \dodoi{10.3847/2041-8213/ac75c3}

\bibitem[{{Murchikova} \& {Witzel}(2021)}]{Murchikova-Witzel2021}
{Murchikova}, L., \& {Witzel}, G. 2021, \apjl, 920, L7, \dodoi{10.3847/2041-8213/ac2308}

\bibitem[{{Narayan} \& {Yi}(1995)}]{Narayan-Yi1995}
{Narayan}, R., \& {Yi}, I. 1995, \apj, 452, 710, \dodoi{10.1086/176343}

\bibitem[{{Nathanail} {et~al.}(2019){Nathanail}, {Porth}, \& {Rezzolla}}]{Nathanail-etal2019}
{Nathanail}, A., {Porth}, O., \& {Rezzolla}, L. 2019, \apjl, 870, L20, \dodoi{10.3847/2041-8213/aaf73a}

\bibitem[{{Novikov} \& {Thorne}(1973)}]{Novikov-Thorne1973}
{Novikov}, I.~D., \& {Thorne}, K.~S. 1973, in Black Holes (Les Astres Occlus), 343--450

\bibitem[{{Okuda}(2014)}]{Okuda2014}
{Okuda}, T. 2014, \mnras, 441, 2354, \dodoi{10.1093/mnras/stu738}

\bibitem[{{Okuda} \& {Das}(2015)}]{Okuda-Das2015}
{Okuda}, T., \& {Das}, S. 2015, \mnras, 453, 147, \dodoi{10.1093/mnras/stv1626}

\bibitem[{{Okuda} \& {Molteni}(2012)}]{Okuda-Molteni2012}
{Okuda}, T., \& {Molteni}, D. 2012, \mnras, 425, 2413, \dodoi{10.1111/j.1365-2966.2012.21571.x}

\bibitem[{{Okuda} {et~al.}(2022){Okuda}, {Singh}, \& {Aktar}}]{Okuda-etal2022}
{Okuda}, T., {Singh}, C.~B., \& {Aktar}, R. 2022, \mnras, 514, 5074, \dodoi{10.1093/mnras/stac1630}

\bibitem[{{Okuda} {et~al.}(2019){Okuda}, {Singh}, {Das}, {Aktar}, {Nandi}, \& {Dal Pino}}]{Okuda-etal2019}
{Okuda}, T., {Singh}, C.~B., {Das}, S., {et~al.} 2019, \pasj, 71, 49, \dodoi{10.1093/pasj/psz021}

\bibitem[{{Olivares} {et~al.}(2019){Olivares}, {Porth}, {Davelaar}, {Most}, {Fromm}, {Mizuno}, {Younsi}, \& {Rezzolla}}]{Olivares-etal2019}
{Olivares}, H., {Porth}, O., {Davelaar}, J., {et~al.} 2019, \aap, 629, A61, \dodoi{10.1051/0004-6361/201935559}

\bibitem[{{Olivares} {et~al.}(2023){Olivares}, {Mo{\'s}cibrodzka}, \& {Porth}}]{Olivares-etal2023}
{Olivares}, H.~R., {Mo{\'s}cibrodzka}, M.~A., \& {Porth}, O. 2023, \aap, 678, A141, \dodoi{10.1051/0004-6361/202346010}

\bibitem[{{Palit} {et~al.}(2019){Palit}, {Janiuk}, \& {Sukova}}]{Palit-etal2019}
{Palit}, I., {Janiuk}, A., \& {Sukova}, P. 2019, \mnras, 487, 755, \dodoi{10.1093/mnras/stz1296}

\bibitem[{{Porth} {et~al.}(2021){Porth}, {Mizuno}, {Younsi}, \& {Fromm}}]{Porth-etal2021}
{Porth}, O., {Mizuno}, Y., {Younsi}, Z., \& {Fromm}, C.~M. 2021, \mnras, 502, 2023, \dodoi{10.1093/mnras/stab163}

\bibitem[{{Porth} {et~al.}(2017){Porth}, {Olivares}, {Mizuno}, {Younsi}, {Rezzolla}, {Moscibrodzka}, {Falcke}, \& {Kramer}}]{Porth-etal2017}
{Porth}, O., {Olivares}, H., {Mizuno}, Y., {et~al.} 2017, Computational Astrophysics and Cosmology, 4, 1, \dodoi{10.1186/s40668-017-0020-2}

\bibitem[{{Porth} {et~al.}(2019){Porth}, {Chatterjee}, {Narayan}, {Gammie}, {Mizuno}, {Anninos}, {Baker}, {Bugli}, {Chan}, {Davelaar}, {Del Zanna}, {Etienne}, {Fragile}, {Kelly}, {Liska}, {Markoff}, {McKinney}, {Mishra}, {Noble}, {Olivares}, {Prather}, {Rezzolla}, {Ryan}, {Stone}, {Tomei}, {White}, {Younsi}, {Akiyama}, {Alberdi}, {Alef}, {Asada}, {Azulay}, {Baczko}, {Ball}, {Balokovi{\'c}}, {Barrett}, {Bintley}, {Blackburn}, {Boland}, {Bouman}, {Bower}, {Bremer}, {Brinkerink}, {Brissenden}, {Britzen}, {Broderick}, {Broguiere}, {Bronzwaer}, {Byun}, {Carlstrom}, {Chael}, {Chatterjee}, {Chen}, {Chen}, {Cho}, {Christian}, {Conway}, {Cordes}, {Geoffrey}, {Crew}, {Cui}, {De Laurentis}, {Deane}, {Dempsey}, {Desvignes}, {Doeleman}, {Eatough}, {Falcke}, {Fish}, {Fomalont}, {Fraga-Encinas}, {Freeman}, {Friberg}, {Fromm}, {G{\'o}mez}, {Galison}, {Garc{\'\i}a}, {Gentaz}, {Georgiev}, {Goddi}, {Gold}, {Gu}, {Gurwell}, {Hada}, {Hecht}, {Hesper}, {Ho}, {Ho}, {Honma}, {Huang}, {Huang}, {Hughes}, {Ikeda}, {Inoue}, {Issaoun},
  {James}, {Jannuzi}, {Janssen}, {Jeter}, {Jiang}, {Johnson}, {Jorstad}, {Jung}, {Karami}, {Karuppusamy}, {Kawashima}, {Keating}, {Kettenis}, {Kim}, {Kim}, {Kim}, {Kino}, {Koay}, {Patrick}, {Koch}, {Koyama}, {Kramer}, {Kramer}, {Krichbaum}, {Kuo}, {Lauer}, {Lee}, {Li}, {Li}, {Lindqvist}, {Liu}, {Liuzzo}, {Lo}, {Lobanov}, {Loinard}, {Lonsdale}, {Lu}, {MacDonald}, {Mao}, {Marrone}, {Marscher}, {Mart{\'\i}-Vidal}, {Matsushita}, {Matthews}, {Medeiros}, {Menten}, {Mizuno}, {Moran}, {Moriyama}, {Moscibrodzka}, {M{\"u}ller}, {Nagai}, {Nagar}, {Nakamura}, {Narayanan}, {Natarajan}, {Neri}, {Ni}, {Noutsos}, {Okino}, {Oyama}, {{\"O}zel}, {Palumbo}, {Patel}, {Pen}, {Pesce}, {Pi{\'e}tu}, {Plambeck}, {PopStefanija}, {Preciado-L{\'o}pez}, {Psaltis}, {Pu}, {Ramakrishnan}, {Rao}, {Rawlings}, {Raymond}, {Ripperda}, {Roelofs}, {Rogers}, {Ros}, {Rose}, {Roshanineshat}, {Rottmann}, {Roy}, {Ruszczyk}, {Rygl}, {S{\'a}nchez}, {S{\'a}nchez-Arguelles}, {Sasada}, {Savolainen}, {Schloerb}, {Schuster}, {Shao}, {Shen}, {Small}, {Sohn},
  {SooHoo}, {Tazaki}, {Tiede}, {Tilanus}, {Titus}, {Toma}, {Torne}, {Trent}, {Trippe}, {Tsuda}, {van Bemmel}, {van Langevelde}, {van Rossum}, {Wagner}, {Wardle}, {Weintroub}, {Wex}, {Wharton}, {Wielgus}, {Wong}, {Wu}, {Young}, {Young}, {Yuan}, {Yuan}, {Zensus}, {Zhao}, {Zhao}, {Zhu}, \& {Event Horizon Telescope Collaboration}}]{Porth-etal2019}
{Porth}, O., {Chatterjee}, K., {Narayan}, R., {et~al.} 2019, \apjs, 243, 26, \dodoi{10.3847/1538-4365/ab29fd}

\bibitem[{{Proga} \& {Begelman}(2003)}]{Proga-Begelman2003}
{Proga}, D., \& {Begelman}, M.~C. 2003, \apj, 582, 69, \dodoi{10.1086/344537}

\bibitem[{{Quataert}(2004)}]{Quataert2004}
{Quataert}, E. 2004, \apj, 613, 322, \dodoi{10.1086/422973}

\bibitem[{{Ressler} {et~al.}(2018){Ressler}, {Quataert}, \& {Stone}}]{Ressler-etal2018}
{Ressler}, S.~M., {Quataert}, E., \& {Stone}, J.~M. 2018, \mnras, 478, 3544, \dodoi{10.1093/mnras/sty1146}

\bibitem[{{Ressler} {et~al.}(2023){Ressler}, {White}, \& {Quataert}}]{Ressler-etal2023}
{Ressler}, S.~M., {White}, C.~J., \& {Quataert}, E. 2023, arXiv e-prints, arXiv:2303.15503, \dodoi{10.48550/arXiv.2303.15503}

\bibitem[{{Ripperda} {et~al.}(2022){Ripperda}, {Liska}, {Chatterjee}, {Musoke}, {Philippov}, {Markoff}, {Tchekhovskoy}, \& {Younsi}}]{Ripperda-etal2022}
{Ripperda}, B., {Liska}, M., {Chatterjee}, K., {et~al.} 2022, \apjl, 924, L32, \dodoi{10.3847/2041-8213/ac46a1}

\bibitem[{{Royster} {et~al.}(2019){Royster}, {Yusef-Zadeh}, {Wardle}, {Kunneriath}, {Cotton}, \& {Roberts}}]{Royster-etal2019}
{Royster}, M.~J., {Yusef-Zadeh}, F., {Wardle}, M., {et~al.} 2019, \apj, 872, 2, \dodoi{10.3847/1538-4357/aafd38}

\bibitem[{{Ryu} {et~al.}(1995){Ryu}, {Brown}, {Ostriker}, \& {Loeb}}]{Ryu-etal1995}
{Ryu}, D., {Brown}, G.~L., {Ostriker}, J.~P., \& {Loeb}, A. 1995, \apj, 452, 364, \dodoi{10.1086/176308}

\bibitem[{{Sano} {et~al.}(2004){Sano}, {Inutsuka}, {Turner}, \& {Stone}}]{Sano-etal2004}
{Sano}, T., {Inutsuka}, S.-i., {Turner}, N.~J., \& {Stone}, J.~M. 2004, \apj, 605, 321, \dodoi{10.1086/382184}

\bibitem[{Scepi {et~al.}(2022)Scepi, Dexter, \& Begelman}]{Scepi-etal2022}
Scepi, N., Dexter, J., \& Begelman, M.~C. 2022, Monthly Notices of the Royal Astronomical Society, 511, 3536, \dodoi{10.1093/mnras/stac337}

\bibitem[{{Shakura} \& {Sunyaev}(1973)}]{Shakura-Sunyaev1973}
{Shakura}, N.~I., \& {Sunyaev}, R.~A. 1973, \aap, 500, 33

\bibitem[{{Singh} {et~al.}(2021){Singh}, {Okuda}, \& {Aktar}}]{Singh-etal2021}
{Singh}, C.~B., {Okuda}, T., \& {Aktar}, R. 2021, arXiv e-prints, arXiv:2101.05474.
\newblock \doarXiv{2101.05474}

\bibitem[{{Sukov{\'a}} {et~al.}(2017){Sukov{\'a}}, {Charzy{\'n}ski}, \& {Janiuk}}]{Sukova-etal2017}
{Sukov{\'a}}, P., {Charzy{\'n}ski}, S., \& {Janiuk}, A. 2017, \mnras, 472, 4327, \dodoi{10.1093/mnras/stx2254}

\bibitem[{Takahashi(2008)}]{Takahashi2008}
Takahashi, R. 2008, Monthly Notices of the Royal Astronomical Society, 383, 1155

\bibitem[{{Tchekhovskoy} {et~al.}(2010){Tchekhovskoy}, {Narayan}, \& {McKinney}}]{Tchekhovskoy-etal2010}
{Tchekhovskoy}, A., {Narayan}, R., \& {McKinney}, J.~C. 2010, \apj, 711, 50, \dodoi{10.1088/0004-637X/711/1/50}

\bibitem[{{Tchekhovskoy} {et~al.}(2011){Tchekhovskoy}, {Narayan}, \& {McKinney}}]{Tchekhovskoy-etal2011}
---. 2011, \mnras, 418, L79, \dodoi{10.1111/j.1745-3933.2011.01147.x}

\bibitem[{{Vourellis} {et~al.}(2019){Vourellis}, {Fendt}, {Qian}, \& {Noble}}]{Vourellis-etal2019}
{Vourellis}, C., {Fendt}, C., {Qian}, Q., \& {Noble}, S.~C. 2019, \apj, 882, 2, \dodoi{10.3847/1538-4357/ab32e2}

\bibitem[{{Wielgus} {et~al.}(2022){Wielgus}, {Marchili}, {Mart{\'\i}-Vidal}, {Keating}, {Ramakrishnan}, {Tiede}, {Fomalont}, {Issaoun}, {Neilsen}, {Nowak}, {Blackburn}, {Gammie}, {Goddi}, {Haggard}, {Lee}, {Moscibrodzka}, {Tetarenko}, {Bower}, {Chan}, {Chatterjee}, {Chesler}, {Dexter}, {Doeleman}, {Georgiev}, {Gurwell}, {Johnson}, {Marrone}, {Mus}, {Psaltis}, {Ripperda}, {Witzel}, {Akiyama}, {Alberdi}, {Alef}, {Algaba}, {Anantua}, {Asada}, {Azulay}, {Bach}, {Baczko}, {Ball}, {Balokovi{\'c}}, {Barrett}, {Baub{\"o}ck}, {Benson}, {Bintley}, {Blundell}, {Boland}, {Bouman}, {Boyce}, {Bremer}, {Brinkerink}, {Brissenden}, {Britzen}, {Broderick}, {Broguiere}, {Bronzwaer}, {Bustamante}, {Byun}, {Carlstrom}, {Ceccobello}, {Chael}, {Chatterjee}, {Chen}, {Chen}, {Cho}, {Christian}, {Conroy}, {Conway}, {Cordes}, {Crawford}, {Crew}, {Cruz-Osorio}, {Cui}, {Davelaar}, {De Laurentis}, {Deane}, {Dempsey}, {Desvignes}, {Dhruv}, {Dzib}, {Eatough}, {Emami}, {Falcke}, {Farah}, {Fish}, {Ford}, {Fraga-Encinas}, {Freeman}, {Friberg},
  {Fromm}, {Fuentes}, {Galison}, {Garc{\'\i}a}, {Gentaz}, {Gold}, {G{\'o}mez-Ruiz}, {G{\'o}mez}, {Gu}, {Hada}, {Haworth}, {Hecht}, {Hesper}, {Ho}, {Ho}, {Honma}, {Huang}, {Huang}, {Hughes}, {Ikeda}, {Impellizzeri}, {Inoue}, {James}, {Jannuzi}, {Janssen}, {Jeter}, {Jiang}, {Jim{\'e}nez-Rosales}, {Jorstad}, {Joshi}, {Jung}, {Karami}, {Karuppusamy}, {Kawashima}, {Kettenis}, {Kim}, {Kim}, {Kim}, {Kim}, {Kino}, {Koay}, {Kocherlakota}, {Kofuji}, {Koch}, {Koyama}, {Kramer}, {Kramer}, {Krichbaum}, {Kuo}, {La Bella}, {Lauer}, {Lee}, {Leung}, {Levis}, {Li}, {Lico}, {Lindahl}, {Lindqvist}, {Lisakov}, {Liu}, {Liu}, {Liuzzo}, {Lo}, {Lobanov}, {Loinard}, {Lonsdale}, {Lu}, {Mao}, {Markoff}, {Marscher}, {Matsushita}, {Matthews}, {Medeiros}, {Menten}, {Michalik}, {Mizuno}, {Mizuno}, {Moran}, {Moriyama}, {M{\"u}ller}, {Musoke}, {Myserlis}, {Nadolski}, {Nagai}, {Nagar}, {Nakamura}, {Narayan}, {Narayanan}, {Natarajan}, {Nathanail}, {Navarro Fuentes}, {Neri}, {Ni}, {Noutsos}, {Oh}, {Okino}, {Olivares}, {Ortiz-Le{\'o}n}, {Oyama},
  {{\"O}zel}, {Palumbo}, {Paraschos}, {Park}, {Parsons}, {Patel}, {Pen}, {Pesce}, {Pi{\'e}tu}, {Plambeck}, {PopStefanija}, {Porth}, {P{\"o}tzl}, {Prather}, {Preciado-L{\'o}pez}, {Pu}, {Rao}, {Rawlings}, {Raymond}, {Rezzolla}, {Ricarte}, {Roelofs}, {Rogers}, {Ros}, {Romero-Canizales}, {Roshanineshat}, {Rottmann}, {Roy}, {Ruiz}, {Ruszczyk}, {Rygl}, {S{\'a}nchez}, {S{\'a}nchez-Arg{\"u}elles}, {S{\'a}nchez-Portal}, {Sasada}, {Satapathy}, {Savolainen}, {Schloerb}, {Schuster}, {Shao}, {Shen}, {Small}, {Won Sohn}, {SooHoo}, {Souccar}, {Sun}, {Tazaki}, {Tilanus}, {Titus}, {Torne}, {Traianou}, {Trent}, {Trippe}, {van Bemmel}, {van Langevelde}, {van Rossum}, {Vos}, {Wagner}, {Ward-Thompson}, {Wardle}, {Weintroub}, {Wex}, {Wharton}, {Wiik}, {Wondrak}, {Wong}, {Wu}, {Yamaguchi}, {Yoon}, {Young}, {Young}, {Younsi}, {Yuan}, {Yuan}, {Zensus}, {Zhang}, {Zhao}, \& {Zhao}}]{Wielgus-etal2022}
{Wielgus}, M., {Marchili}, N., {Mart{\'\i}-Vidal}, I., {et~al.} 2022, \apjl, 930, L19, \dodoi{10.3847/2041-8213/ac6428}

\bibitem[{{Yusef-Zadeh} {et~al.}(2020){Yusef-Zadeh}, {Royster}, {Wardle}, {Cotton}, {Kunneriath}, {Heywood}, \& {Michail}}]{Yusef-Zadeh-etal2020}
{Yusef-Zadeh}, F., {Royster}, M., {Wardle}, M., {et~al.} 2020, \mnras, 499, 3909, \dodoi{10.1093/mnras/staa2399}

\end{thebibliography}
\bibliographystyle{aasjournal}


\label{lastpage}
\end{document}